\documentclass[10pt,journal,compsoc]{IEEEtran}
\ifCLASSOPTIONcompsoc
  \usepackage[nocompress]{cite}
\else
  \usepackage{cite}
\fi

\ifCLASSINFOpdf
\else
\fi

\usepackage{bbding}
\usepackage{amsfonts}
\usepackage{amsmath}
\usepackage{graphicx}
\usepackage{float}
\usepackage{subfigure}
\usepackage{bm}
\usepackage{courier}
\usepackage{booktabs}
\usepackage{multirow}

\usepackage{rotating}
\usepackage{amsthm}
\usepackage{footnote}
\usepackage{color}
\usepackage{stfloats}
\usepackage{threeparttable}
\usepackage{adjustbox}

\usepackage{orcidlink}
\usepackage{colortbl, booktabs}
\usepackage{pifont}
\usepackage{textcomp}
\usepackage{stfloats}
\usepackage{url}
\usepackage{verbatim}
\usepackage{graphicx}
\usepackage{algorithm}
\usepackage{array}
\usepackage{amsmath,amsfonts}
\hypersetup{hidelinks}
\usepackage[noend]{algorithmic}
\usepackage{booktabs}
\usepackage{xcolor}
\usepackage{enumitem}
\usepackage{hyperref}
\hypersetup{
	colorlinks=true,
	citecolor=green,
	linkcolor=red,
}

\usepackage[most]{tcolorbox}
\usepackage{amssymb}

\begin{document}

\title{Detecting Malicious Concepts Without Image Generation in AI-Generated Content (AIGC)}

\author{Kun Xu$^{\orcidlink{0000-0002-1866-4433}}$, Wenying Wen$^{\orcidlink{0000-0002-3098-4640}}$,~\IEEEmembership{Member,~IEEE}, Shuren Qi$^{\orcidlink{0000-0003-0574-2313}}$, Tao Wang$^{\orcidlink{0000-0001-5532-3999}}$, Yushu Zhang$^{\orcidlink{0000-0001-8183-8435}}$,~\IEEEmembership{Senior Member,~IEEE}, Yuming Fang$^{\orcidlink{0000-0002-6946-3586}}$,~\IEEEmembership{Fellow,~IEEE}

\thanks{This work was supported in part by the National Natural Science Foundation of China under Grant 62522112, the Ganpo Talent Program of Jiangxi Province under Grant gpyc20240012, the Outstanding Youth Fund Program of Jiangxi Province under Grant 20252BAC220008, the Jiangxi Key Research and Development Program under Grant 20261BCE310050, and the China Scholarship Council Program under Grant 202506830129. Corresponding author: Wenying Wen (e-mail: wenyingwen@sina.cn).}

\thanks{
Kun Xu and Tao Wang are with the College of Computer Science and Technology, Nanjing University of Aeronautics and Astronautics, Nanjing 210016, China (e-mail: xukun930@nuaa.edu.cn; wangtao21@nuaa.edu.cn).

Wenying Wen, Yushu Zhang, and Yuming Fang are with the School of Computing and Artificial Intelligence, Jiangxi University of Finance and Economics, Nanchang 330044, China, and also with the Jiangxi Provincial Key Laboratory of Multimedia Intelligent Processing, Nanchang 330044, China (e-mail: wenyingwen@sina.cn; yushu@nuaa.edu.cn; fa0001ng@e.ntu.edu.sg).

Shuren Qi is with the Department of Data Science, City University of Hong Kong, Hong Kong, China (e-mail: shurenqi@cityu.edu.hk).
}
}


\markboth{K. Xu \MakeLowercase{\textit{et al.}}: Detecting Malicious Concepts Without Image Generation in AI-generated content (AIGC)} {K. Xu \MakeLowercase{\textit{et al.}}: Detecting Malicious Concepts Without Image Generation in AI-generated content (AIGC)}

\IEEEtitleabstractindextext{%
\begin{abstract}
The task of text-to-image generation has achieved tremendous success in practice, with emerging concept generation models capable of producing highly personalized and customized content. Fervor for concept generation is increasing rapidly among users, and platforms for concept sharing have sprung up. The concept owners may upload malicious concepts and disguise them with non-malicious text descriptions and example images to deceive users into downloading and generating malicious content. The platform needs a quick method to determine whether a concept is malicious to prevent the spread of malicious concepts. However, simply relying on concept image generation to judge whether a concept is malicious requires time and computational resources. Especially, as the number of concepts uploaded and downloaded on the platform continues to increase, this approach becomes impractical and poses a risk of generating malicious content. In this paper, we propose Concept QuickLook, the first systematic work to incorporate malicious concept detection into research, which performs detection based solely on concept files without generating any images. We define malicious concepts and design two operational modes for detection: concept matching and fuzzy detection. Extensive experiments demonstrate that the proposed Concept QuickLook can detect malicious concepts and demonstrate practicality in concept sharing platforms. We also design robustness experiments to further validate the effectiveness of the solution. We hope this work can initiate malicious concept detection tasks and provide some inspiration.

\end{abstract}

}

\maketitle

\IEEEdisplaynontitleabstractindextext

%
\IEEEpeerreviewmaketitle

\IEEEraisesectionheading{\section{Introduction}\label{sec:intro}}

%
%
%
%

\IEEEPARstart{T}{here} have been immense advances in generative models that use text as an input condition and control the direction of generation in recent years, especially the text-to-image (T2I) generation models like Stable Diffusion (SD) \cite{sdpaper}. Diffusion models are extremely realistic in generating images that give a detailed description through text \cite{ramesh2022}, \cite{NEURIPS2022_ec795aea}, \cite{bar2023multidiffusion}, \cite{10841434}. With the rapid development of these generative models, there is a demand for the model-generated content to have personalization and customization \cite{ma2023subject, shi2023instantbooth}. This requires the models to be subject-driven as in Textual Inversion \cite{gal2022image} and Dreambooth \cite{Ruiz_2023_CVPR}, as opposed to the normal text-image generation models. Currently, expanding on the above generation models is defined as \textit{concept generation model} \cite{liu2023cones}, \cite{Kumari_2023_CVPR}, \cite{liu2023cones2}, \cite{li2024blip}, \cite{10.1145/3659578}, which abstracts the subject to be personalized into a concept that makes the model generate content that is difficult to describe in text.

\textbf{Generating and Applying Concepts.} The workflow of the concept generation model based on SD can be divided into two parts: \textit{generating concepts} and \textit{applying concepts}. The former is the process of extracting the concept from several image examples of the concept \cite{gal2022image, Ruiz_2023_CVPR, 10489849}, while the latter focuses on how to create new content from the extracted concept \cite{ma2023subject, safaee2023clic}. With the development of related community platforms such as the \textit{Civitai} and \textit{Hugging Face}, concept owners can easily upload their extracted concept files and share them with other users for creation.

\textbf{Understandable and Non-understandable Concepts.} During the process of uploading a concept to the platform by the owner, the concept file cannot be visualized to the user by itself because it is an embedding consisting of one or more vectors. Usually, it is necessary to attract users to download and utilize it in the form of descriptive texts and example diagrams. The users build up a ``concept'' of an object in their brains through the example diagrams and text descriptions presented, which are described in this paper as \textit{user understandable concepts} (UUC). In addition, users will make a natural correspondence between this ``concept'' and the concept files provided by the platform for download. Without example diagrams and text descriptions, it would be difficult to understand and differentiate between the concepts, and this paper describes them as \textit{user non-understandable concepts} (UNC).
\begin{figure*}[h]
    \centering
    \includegraphics[width=\linewidth]{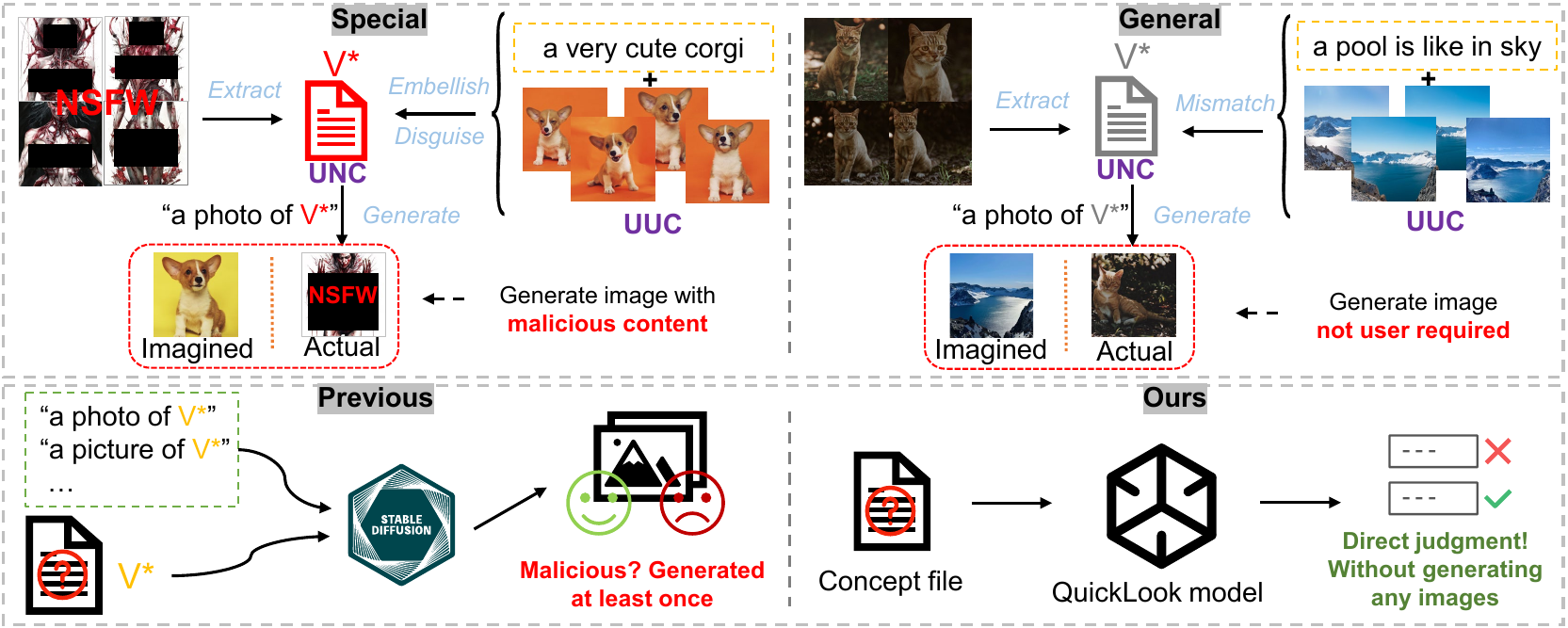}
    \caption{Overview. \textit{\textbf{Top}}: The left is the special case, where the actual concept file is malicious, but it is presented in a harmless form after disguise and embellishment, it will generate harmful content. The right is the general case, where the actual concept file mismatches the concept descriptions, generating images that are not user required. \textit{\textbf{Bottom}}: The left shows the inefficient method of determining by generating images at least once. On the right is Concept QuickLook, which achieve directly judge without generating any images}.
    \label{figure1-overview}
\end{figure*}

\textbf{Security Risks and Mitigation Strategies.} Concept generation models are developing rapidly, there is a widespread concern about the security risks involved in both the processes of generating concepts and applying concepts \cite{Gandikota_2024_WACV}, \cite{Van_Le_2023_ICCV}, \cite{zhang2023backdooring}, \cite{Kumari_2023_ICCV}, \cite{lyu2023onedimensional}. Some security issues are gradually exposed \cite{Degeneration-Tuning}, \cite{feng2023catch}, \cite{tsai2023ring}, \cite{11316185}, \cite{kim2025trainingfree}. While extracting concepts, on one hand, malicious concepts that violate ethics and laws may be extracted, coupled with the owner's malicious disguise as innocuous concepts uploaded to the platform. On the other hand, users are confused by normal text descriptions and example diagrams when applying concepts and are prone to generate images containing malicious content. A security concern regarding the concept is illustrated in Figure \ref{figure1-overview} (\textit{Top}). We explain the implementation logic for disguising and embellishing malicious concepts, as well as the malicious mismatch of concepts.

Existing moderation in concept-sharing scenarios is largely reactive: in practice, a concept is often judged only after at least one image has been generated. This is costly, slow, and may itself expose users or reviewers to harmful outputs \cite{11395536}. Therefore, there is a practical need for an upload-time safety mechanism that can assess concept risk directly from the concept file before any image generation takes place.

\textbf{Our Work.} We assume a scenario in which the owner is malicious, extracts the offending concepts, and generates a malicious concept file that is then uploaded to the platform. The user downloads the malicious concept file for generating personalized images based on misleading text descriptions and example diagrams, but the user cannot effectively determine whether the concept file generates harmful content before the image is generated. Moreover, it may take more than one generation to determine whether a concept file is malicious. This can be potentially jeopardized for the user, and also significantly undermines the rights and credibility of the platform. It is necessary to detect malicious concepts, i.e., the platform side needs to provide users with a way to distinguish between them. To address these issues, we propose the \textbf{Concept QuickLook}.

Concept QuickLook is a solution that informs the user's judgment regarding the potential risks of a concept file (Figure \ref{figure1-overview} (\textit{Bottom})). It also assesses whether the generated image may contain malicious content before the user utilizes the concept file for personalized creative generation. Our work is the first to clearly define malicious concepts and propose targeted detection solutions. We define two cases of the malicious concepts: \textit{special} and \textit{general}.
(\romannumeral 1) \textit{The first case involves concepts that are inherently malicious and generate harmful content, but are disguised and embellished}.
(\romannumeral 2) \textit{The second case involves concepts that are maliciously mismatched with their text descriptions and example diagrams, preventing the generation of the desired content for the user}. Concept QuickLook employs two detection workflows: one that detects if the concept itself is malicious by combining the concept's description to provide a judgment, and another that judges whether the concept belongs to a known concept class without relying on the description. To evaluate the performance of Concept QuickLook, we conducted extensive experiments, covering model training, detection performance of the two detection modes, and robustness assessment. We carried out targeted case studies for both special and general malicious concepts. In addition to using quantitative metrics for evaluation, we also introduced human evaluation methods to better align with user needs. The experimental results demonstrate the effectiveness of our work. The major contributions of this paper are summarized as follows:
\begin{itemize}[leftmargin=*]
\item We are the first to define the malicious concept in the concept sharing process, and propose a solution called Concept QuickLook to achieve quickly malicious concept detection.

\item We analyze the generation mechanism of the concept generation model and the whole process of concept file sharing. The embedding vectors in the concept file are found to be the primary factor controlling the content of the generated subjects, and it can serve to detect whether the personalized generated content is malicious or not.

\item We design two operational modes for the QuickLook model: concept matching and fuzzy detection. This paper demonstrates that these two modes effectively meet the requirements for malicious concept detection in the current concept sharing platform scenario.

\item We conducted extensive experiments, including effectiveness evaluation, baseline comparisons, manual scoring, and robustness testing. The results demonstrate that our proposed method implements the ability to distinguish malicious concepts without even one generation, which effectively protects concept sharing platforms and users.
\end{itemize}

\section{Related Works}

We briefly review the closely related text-to-image generation models, generating concepts and applying concepts.

\textbf{Text-to-image Generation Models.} AI-generated content (AIGC) \cite{cao2023comprehensive}, \cite{xu2024unleashing}, \cite{wang2023security}, \cite{10230895} has experienced significant expansion in recent years. Image generation has evolved from the pursuit of high resolution and more realism to a more flexible and broadly applicable to a variety of tasks, which allows for customized generation to fit user needs. GAN-based (Generative Adversarial Network \cite{goodfellow2014generative}) generation models have previously achieved better generation quality, and the images generated by GAN are usually highly realistic and lifelike \cite{10.1145/3439723}. GAN-INT-CLS \cite{reed2016generative} was the first to realize GAN-based text-to-image generation, and many subsequent works \cite{Cheng9156682}, \cite{Huang2021Unifying}, \cite{Ruan9710042} have demonstrated its powerful generative capabilities. However, GAN faces limitations such as poor performance during complex scenes and unstable training.

With diffusion probabilistic model (DM) \cite{pmlr-v37-sohl-dickstein15}, denoising diffusion probabilistic model (DDPM) \cite{NEURIPS2020_4c5bcfec}, latent diffusion model (LDM) \cite{sdpaper} and further development to SD, diffusion models have received much attention in image generation due to their powerful capabilities \cite{Croitoru10081412, Yang2023Diffusion}. An OpenAI study shows that diffusion models beat GANs in image synthesis \cite{NEURIPS2021_49ad23d1}. VQ-Diffusion \cite{Gu9879180} is a T2I generation model based on a vector quantized variational autoencoder (VQ-VAE) with a latent space modeled by a conditional variant of DDPM, which can handle more complex scenes and dramatically improve image quality. Imagen \cite{NEURIPS2022_ec795aea} achieves remarkable photorealism and a profound level of language understanding, with unprecedented realism and ensures precise text alignment. InstaFlow \cite{liu2024instaflow} is a proposed novel T2I generation model to turn SD into an ultra-fast text-conditioned pipeline. A recent research work HIVE \cite{zhang2023hive} attempts to combine T2I modeling with human feedback to control the output of generated images through editorial commands. Further, a comprehensive benchmark for open-world compositional T2I generation, T2I-CompBench \cite{NEURIPS2023_f8ad010c}, has been proposed.

\textbf{Generating Concepts.} Generating concepts is often described as a process where multiple images of shared concepts are used for concept extraction through well-designed models, such as Textual Inversion \cite{gal2022image}, DreamBooth \cite{Ruiz_2023_CVPR} and Lora-DreamBooth \cite{hu2022lora}. It is considered as an extension of the T2I generation task. The generative concepts technique is first directed toward extracting the concept of an entity \cite{10.1145/3618315}, e.g., a person, an automobile, a ragdoll toy, and so on. Cones \cite{liu2023cones, liu2023cones2}, Break-A-Scene \cite{Avrahami2023Break} and Custom Diffusion \cite{Kumari_2023_CVPR} implement multi-concept extraction. CatVersion \cite{zhao2023catversion} learns the personalization concept by concatenating embeddings on the feature-dense space of a text encoder in the diffusion model. For the extraction of style concepts, StyleDrop \cite{NEURIPS2023_d33b177b} is a method that faithfully follows a specific style and effectively learns new ones. InST \cite{Zhang_2023_CVPR} perceives styles as learnable text descriptions of paintings, which can efficiently and accurately learn key artistic style information about an image. Specialist Diffusion \cite{Lu_2023_CVPR} is a plug-and-play framework for style concept extraction with further performance improvements.

\textbf{Applying Concepts.} Those working on how to better extract concepts usually demonstrate the excellent generative power of their methods. CLiC \cite{safaee2023clic} performs contextual concept learning on the concept within the inclusion mask and the surrounding image region to acquire local visual concept. Subject-Diffusion \cite{ma2023subject} is a novel open-domain personalized image generation model that requires only a single reference image to support single-subject or multi-subject personalization generation with no need for test-time fine-tuning. Moreover, one kind of application phase of concepts can be summarized as concept assisting. Anti-DreamBooth \cite{Van_Le_2023_ICCV} destroys the generation quality of any DreamBooth model trained against malicious use by adding perturbations to the images. Concept censorship \cite{zhang2023backdooring} proposes to inject backdoors into Textual Inversion embeddings and select some sensitive words as triggers during training so that the models do not generate malicious images. Recent studies have further examined security risks and defenses for customized diffusion systems, including text-encoder backdoor defense \cite{11112743} and dynamic-attention-based backdoor detection \cite{11300728}. Watermarking techniques are also orthogonal, as they aim at ownership verification rather than content safety inspection \cite{Li_2025_ICCV}. Concept watermarking \cite{feng2023catch} provides a proactive adversarial defense approach by embedding identifiable patterns into concept embeddings to prevent unauthorized use or tampering. While watermarking focuses on provenance verification and content traceability, our Concept QuickLook complements it by providing post-upload malicious concept detection without requiring image generation.

\section{Problem Formulation}
In this section, we first introduce the preliminary work, which mainly includes the implementation logic of concept extraction and the introduction of the three roles involved in the concept sharing process. Next, we define malicious concepts and present the special and general cases of malicious concepts. Finally, we introduce the threat model.

\subsection{Preliminary}
LDM utilizes an image encoder $\varphi$ to convert images $x\in \mathbb{R}^{H \times W \times 3} $ into a spatial latent code $z=\varphi(x)$. The decoder $D_e$ learns to map such latents back to images. A conditional denoising autoencoder $\epsilon_\theta(z_t,t,c_\theta(y))$ generates images given a specific text $y$ as a condition. $c_\theta(y)$ is a model that maps a conditioning input $y$ into a conditioning vector. The loss of conditional LDM is designed as:
\begin{equation}\label{eq1}
L_{LDM}:=\mathbb{E}_{z\sim\varphi(x),y,\epsilon\sim\mathcal{N}(0,1),t}\Big[\|\epsilon-\epsilon_\theta(z_t,t,c_\theta(y))\|_2^2\Big],
\end{equation}
where the time steps $t \in \{ 1,...,\mathcal{T} \}$, $z_t$ is the latent noised to time $t$, $\epsilon$ is the unscaled noise sample from the Gaussian distribution $\mathcal{N}(0,1)$.

The goal of Textual Inversion is to convert new concepts into pseudo-words in the text encoder embedding. For Textual Inversion, the direct optimization of the LDM loss is used:
\begin{equation}\label{eq2}
e^{*}=\mathop{\arg\min}_{e} \mathbb{E}_{z\sim \varphi(x),y,\epsilon\sim \mathcal{N}(0,1),t}{\Big[}||\epsilon-\epsilon_{\theta}(z_{t},t,c_{\theta}(y))||_{2}^{2}{\Big]},
\end{equation}
where $e^{*}$ is the embedding vector. $e^{*}$ is found through direct optimization by minimizing the LDM loss. In conducting general and intuitive conditional text editing, we designate a placeholder string $V^{*}$ (pseudo-word) to represent the newly learned concept. Table \ref{tab:ND} lists the notations and their definitions for this paper.
\begin{table}
    \centering
    \caption{Notations and Definitions}
    \scalebox{1}{
    \begin{tabular}{|l||l|}
        \hline
 		Notation & Definition \\
        \hline
            $C, C_n, C_m$ & The concept, the normal/malicious concept \\
            $C_f$ & The concept file \\
		$C_T$ & The set of confirmed concept \\
            $e, e^c$ & The concept embedding vector \\
            $\mathcal{R}_{cls}$ & The concept class \\
            $\mathcal{E}$ & The embedding space \\
		$\epsilon_\theta$ & The conditional denoising autoencoder \\ 
		$T_d, E_d$ & The text description, the example diagram  \\
		$C_{f_u}, C_{f_n}, C_{f_m}$ & The unknown/normal/malicious concept file \\
		$V^{*}$ & The pseudo-word \\
		$\mathcal{D}$ & The training set \\
		$\sigma$ & The QuickLook model \\
            $\Gamma _{ext}$ & The concept extraction model \\
            $\rm{Q}$ & The configuration information in concept file \\
            \hline
		$\mathcal{O}$ & The concept owner \\
		$\mathcal{P}$ & The concept sharing platform \\
		$\mathcal{U}$ & The user of concept sharing platform \\
     \hline
    \end{tabular}
    }
    \label{tab:ND}
\end{table}

\begin{figure}[t]
    \centering
    \includegraphics[width=\linewidth]{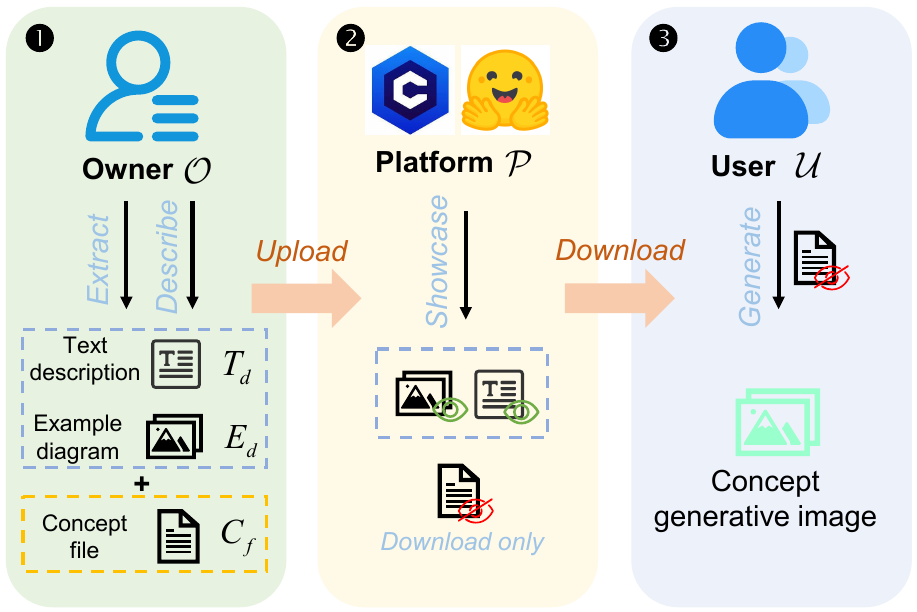}
    \caption{Introduction to the three roles of owner, platform and user in concept sharing.}
    \label{figure2-roles}
\end{figure}

The embedding space $\mathcal{E}$ is a high-dimensional vector space where each vector represents a specific concept, text description, or image feature \cite{pmlr-v139-radford21a, gal2022image}. By operating within this space, connections and transformations between different modalities can be achieved. We align the characterization with practical scenarios, and the interaction process of the three parties involved in concept sharing is illustrated in Figure \ref{figure2-roles}. We show a scenario where a concept owner uploads the concept file with a text description and example diagram to the platform, and a user downloads the concept file in interest to local area and generates a concept generative image. The three roles shown work together to advance the sharing and exchange of concepts \cite{feng2023catch}.

Our detection objective is to determine whether an uploaded concept file is likely to produce harmful content or is semantically inconsistent with its declared description, rather than to predict every pixel of a future generated image. For concept generation methods such as Textual Inversion, the uploaded file mainly contributes the learned concept embedding $e^c$, which is the compact carrier of subject semantics injected into the text encoder. Therefore, for the platform-side pre-screening task considered in this paper, analyzing the embedding is sufficient to capture the semantic signal that is most directly controlled by the uploaded concept file.

\textbf{Role} \ding{182}: the concept owner $\mathcal{O}$. We assume that the owner has a certain level of technical proficiency, enabling them to extract concepts locally, upload and share concept files, and provide the text descriptions $T_d$ and example images $E_d$ for the concepts. The process of concept extraction can be formulated as:
\begin{equation}\label{eq-1}
e^c = {\Gamma _{ext}}(\varphi ({\rm{\mathbf{x}}})),{\rm{\mathbf{x}}} = \{ ({x_1},{x_2},...,{x_i}) \}^c ,
\end{equation}
where $\Gamma_{ext}$ is the concept extraction model, optimized through Equation (\ref{eq2}). The uploaded concept file can be represented as ${C_f} = \{ ({e^c},\rm{Q})\}$, where $e^c \in \mathcal{E}$ denotes the learned concept embedding and $\rm{Q}$ denotes the auxiliary configuration information stored in the file. In our detection pipeline, $C_f$ is first parsed into $e^c$, and all subsequent judgments are performed in the embedding space $\mathcal{E}$.

\textbf{Role} \ding{183}: the concept sharing platform $\mathcal{P}$. We assume that the platform to receive the owner's concept file $C_f$, as well as text descriptions $T_d$ and example images $E_d$. The task of the platform is to make $T_d$ and $E_d$ (corresponding to UUC) available for showcase, and the corresponding $C_f$ (corresponding to UNC) available for download, in addition to providing pseudo-word $V^{*}$. Therefore, a concept can be expressed as $C \leftarrow ({C_f},{T_d},{E_d})$.

\textbf{Role} \ding{184}: the concept sharing platform user $\mathcal{U}$. We assume that the user browses concepts on the sharing platform by reviewing text descriptions $T_d$ and example images $E_d$, downloads concept file $C_f$ to their local devices, utilizes them for personalized concept image generation.

\subsection{Malicious Concept Definition}\label{sec-mcd}
We mentioned that concepts are sharing on the platform in the form of concept files (``.pt'' and ``.safetensors'' files). We are aware that files may contain malicious scripts, etc., but this is not the case for the malicious concepts defined. Concept files $\{C_f\}=\{C_{f_n}, C_{f_m} \}$, where $C_{f_n}$ is normal, $C_{f_m}$ is malicious. The pseudo-word represents a ``concept'' that is subjectively portrayed by the user through $T_d$ and $E_d$. The normal concept can be expressed as $C_n \leftarrow (C_{f_n}^c,{T_d^c},{E_d^c})$, and the malicious concept can be expressed as $C_m \leftarrow (C_{f_m}^{cm},{T_d^c},{E_d^c})$, where the superscript $cm$ indicates association with the malicious concept.

We are concerned with whether the concept files produce malicious or incorrect content when they are used for personalized and customized image generation. Therefore, we establish two cases for the definition of malicious concepts.
\begin{tcolorbox}
	[breakable,		                    
	arc=0mm, auto outer arc,            
	boxrule= 0pt,                        
	boxsep = 0mm,                       
	left = 1mm, right = 1mm, top = 1mm, bottom = 1mm, 
	]
	{\textbf{Definition Case 1: \textit{Special Malicious Concept}.} Concepts that directly contain or reproduce NSFW content (e.g., pornography, violence, or gore), possibly disguised or stylized before being uploaded to the concept-sharing platform.}
 \end{tcolorbox}

Special Malicious Concepts are characterized by embedding-level proximity to harmful semantic prototypes, even when stylized or disguised. Let $\mathfrak{P}=\{\mathfrak{p}_1, \mathfrak{p}_2, \dots, \mathfrak{p}_K\}$ be a curated set of prototype embeddings representing well-defined harmful classes (e.g., sexual content, graphic violence). We define Malicious-Prototype Similarity (MPS) as follows.
\begin{equation}
\text{MPS}(e^c) \triangleq \max_{\mathfrak{p} \in \mathfrak{P}} \cos(e^c, \mathfrak{p}),
\end{equation}
where $\cos(\cdot,\cdot)$ denotes cosine similarity. A concept is flagged as a Special Malicious Concept when $\text{MPS}(e^c) \ge \tau_{\text{MPS}}$, where $\tau_{\text{MPS}}$ is selected on a held-out validation set by maximizing the F1-score of harmful-concept detection.

\begin{tcolorbox}
	[breakable,		                    
	arc=0mm, auto outer arc,            
	boxrule= 0pt,                        
	boxsep = 0mm,                       
	left = 1mm, right = 1mm, top = 1mm, bottom = 1mm, 
	]
	{\textbf{Definition Case 2: \textit{General Malicious Concept}.} Concepts that are inherently benign but are maliciously mislabeled or mismatched with misleading text descriptions or example images during uploading, leading to deceptive or harmful outputs.}
 \end{tcolorbox}

General Malicious Concepts are not harmful due to their intrinsic semantics, but rather because the uploaded concept is intentionally mislabeled or assigned to an incorrect class.  To make this case operational, we quantify whether a concept embedding is statistically inconsistent with the distribution of the class it claims to belong to. For each legitimate class $y$, we compute from benign datasets its embedding mean $\mu_y$ and covariance matrix $\Sigma_y$. We define the Embedding Distribution Deviation Score (EDS) as the Mahalanobis distance:
\begin{equation}
\text{EDS}(e^c, y)
\triangleq
\sqrt{(e^c - \mu_y)^{\top} \Sigma_y^{-1} (e^c - \mu_y)}.
\end{equation}
A concept is flagged as a General Malicious Concept when its deviation exceeds a calibrated threshold $\mathrm{EDS}(e^c,y)\ge \tau_{\text{EDS}}$, where $\tau_{\text{EDS}}$ is selected on a held-out validation set to balance false positives and false negatives.
\begin{figure}
    \centering
    \includegraphics[width=\linewidth]{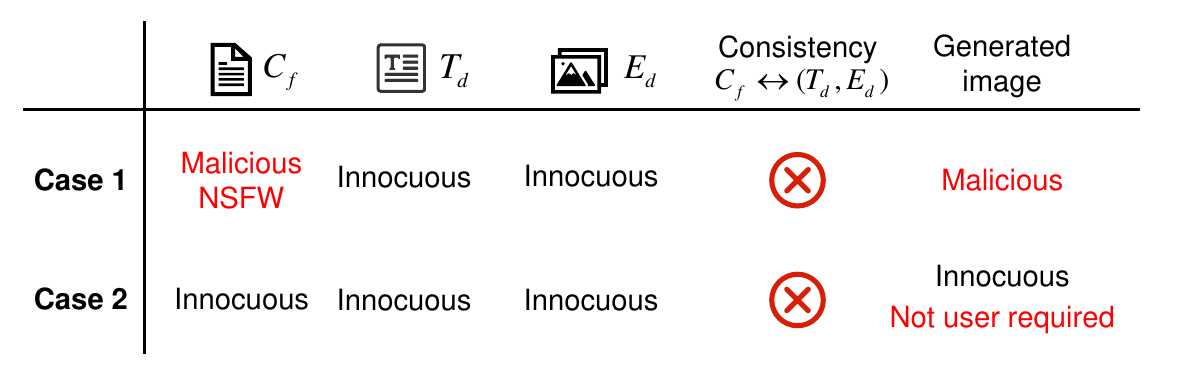}
    \caption{Comparison of two cases for the malicious concept.}
    \label{figure3-cases}
\end{figure}

The comparison of the two cases of malicious concept generation as set in our framework is shown in Figure \ref{figure3-cases}. It should be noted that in Case 1, the concept extracted itself is malicious, and the owner embellishes their concept with innocuous $T_d$ and $E_d$. This case of malicious concepts emerge from the process of concept extraction. As for Case 2, it represents a broader definition of malicious concepts, where users cannot generate the required images. This approach is cost-effective and more widespread. This type of malicious concepts emerge after the process of concept extraction, where the owner maliciously mismatches the concept with $T_d$ and $E_d$.

There has been extensive discussion about the SD safe filter on enthusiast forums, with some users complaining about false triggers and considering it a nuisance. In these forums, users also share methods to bypass these safety checks, and our experiments confirm that such checks can indeed be circumvented. Additionally, the malicious concepts defined in Case 2 are inherently difficult for the safe filter to detect, as they do not contain NSFW content. It is important to note that malicious concept detection is fundamentally different from prompt injection or model misuse detection, as it focuses on the semantic properties of the uploaded concept embedding itself, independent of user prompts or image generation outputs. For Case 1, the key question is whether the uploaded concept embedding is semantically close to harmful prototypes; this is naturally an embedding-space property. For Case 2, the key question is whether the uploaded embedding is compatible with the claimed concept class or declared description; this is also a semantic consistency problem in embedding space. Hence, both cases can be formulated as concept-semantic judgments on $e^c$, without requiring image rendering.
\begin{figure*}[t]
    \centering
    \includegraphics[width=\linewidth]{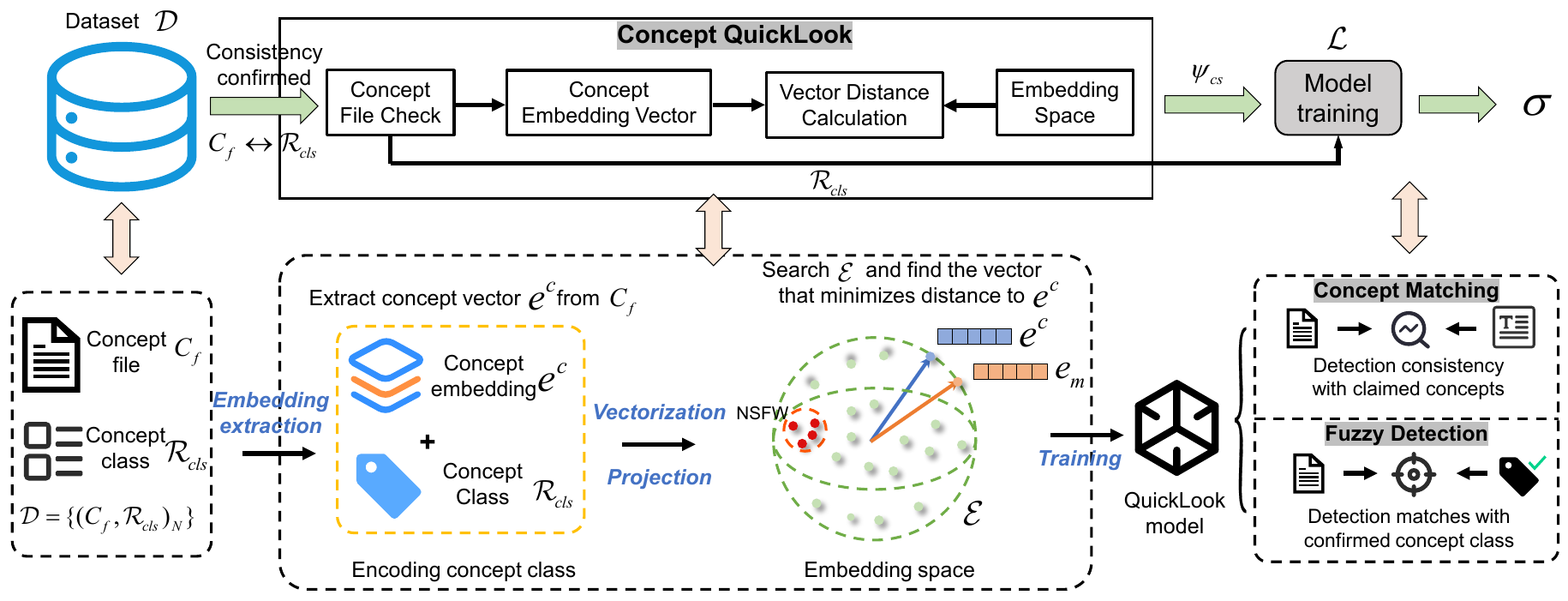}
    \caption{Outline of the Concept QuickLook. The top part of the illustration presents the Concept QuickLook workflow and the QuickLook model training process. The bottom part provides a detailed and intuitive description.}
    \label{figure4-outline}
\end{figure*}

\subsection{Threat Model}\label{sec-tm}
In this work, we focus on the scenario of malicious concept detection. This is a completely new endeavor. 

\textbf{Malicious Concept Owner's Capabilities and Objectives.} A concept upload is abstracted into three parts, whose constituent datasets can be denoted as $ \{ (C_f, T_d, E_d) \} $. In $C_f$ carrying the concept information is the embedding vector $e^c$. In this paper, we consider the case where the concept owner is malicious, i.e., they upload a file containing a malicious concept and mislead the platform and users through normal text descriptions $T_d$ and example diagrams $E_d$. Additionally, another general type of malicious concept involves the owner maliciously mismatching the concept file $C_f$ with $T_d$ and $E_d$, preventing users from generating the required concept content. In detail, since the concept file is not directly observable for its content, the malicious owner extracts the concept and disguises it as a normal upload. Malicious concept owners share disguised concepts, causing users to generate images containing malicious content, thereby undermining the credibility of the concept sharing platform, eroding user trust in the platform.

\textbf{Concept Sharing Platform's Capabilities and Requirements.} The challenge previously was the lack of a definition for what constitutes malicious concepts and the absence of a solution for detecting them. Users could only judge whether a concept was malicious by using the generated images. We assume that the platform is trustworthy and committed to maintaining its reputation by defending the rights of all parties. The platform would like to pre-determine whether an uploaded concept file will generate malicious content, and accordingly provide risk alerts to users who need to download it or simply delete the concept file. We assume that the platform needs to automatically perform malicious concept detection tasks, which cannot be quickly achieved through image generation from concept embeddings. The platform requires a technical solution to rapidly complete the detection of concept files with minimal resource consumption. From a deployment perspective, Concept QuickLook can be integrated as an upload-time pre-screening module before a concept is published on the platform. After parsing the uploaded concept file into its embedding, the platform can apply concept matching and fuzzy detection to produce a risk signal, and then route flagged or borderline cases to manual review while allowing benign concepts to proceed normally. This design keeps image generation outside the moderation loop, reducing both computational overhead and unnecessary exposure to harmful content. Moreover, we attempted to contact the support teams of the current mainstream concept sharing platforms (\textit{Civitai} and \textit{Hugging Face}) to express our concerns about the potential spread of malicious concepts on their platforms and to inquire about existing solutions. However, we did not receive any response.

\textbf{Adversarial Evasion Scenarios.} In addition to malicious concept authors, we also consider the possibility of adversarial users who attempt to evade detection by perturbing concept embeddings. Formally, an adversarial example can be expressed as
\begin{equation}
e' = e + \delta, \quad \|\delta\|_p \le \epsilon,
\end{equation}
where $e$ is the original embedding, $\delta$ is a small perturbation crafted to alter the detection outcome, and $\epsilon$ controls the perturbation budget. Such embedding-level evasion does not modify the concept file semantically but aims to mislead detection by altering distances in the embedding space. In this work, we treat such perturbations as a stress-test setting for the detector rather than a full adaptive attack model. The purpose is to examine whether small embedding-space perturbations can change the detector's semantic judgment, while keeping the underlying concept semantics approximately unchanged.

\section{Concept QuickLook}
\subsection{QuickLook Model Implementation}\label{sec-qmi}
Our data acquisition methods are mainly divided into: local training \cite{gal2022image} and platform downloads (\textit{Civitai} and \textit{Hugging Face}). We perform content generation checks on the downloaded concept files and label all data with concept classes, symbolically denoted as $\mathcal{D}=\{ (C_f, \mathcal{R}_{cls})_N \}$. Additionally, it should be noted that the number of embedding vectors for concepts during the extraction process is a parameter that is set manually. We analyzed many concept files on the concept sharing platform and found that the number of vectors is indeed a variable value. We symbolically represent it as:
\begin{equation}\label{eq4}
C \to {C_f} = ( e^c_n,\rm{Q}) , n \in {\mathbb{Z}^ {+}},
\end{equation}
where $e^c_n$ represents the embedding of $n$ vectors for the concept. The outline of Concept QuickLook is illustrated in Figure \ref{figure4-outline}. The core functionality of Concept QuickLook relies on the QuickLook model $\sigma $ and the embedding vector-concept class pairs $\langle {e^c} \odot {{\cal R}_{cls}}\rangle$, as shown in Equation (\ref{eq-map}). In practice, the embedding vector-concept class pairs are derived from our constructed dataset $\mathcal{D}$. Algorithm \ref{algo_QL} shows the QuickLook model training. The $\sigma$ learns these pairs through training, enabling it to output the corresponding concept class when given an input embedding vector. This is fundamentally different from nearest-neighbor search. NNS retrieves the most similar stored embedding based only on pairwise distance, whereas QuickLook learns class-level semantic structure from the training set and outputs a class-aware prediction for a given concept embedding. Specifically, the actual composition of dataset $\mathcal{D}$ consists of concept files $C_f$ and labeled concepts class $\mathcal{R}_{cls}$. It is necessary to extract cf and convert it into embedding vectors $e^c$, and then pair $e^c$ with $\mathcal{R}_{cls}$,
\begin{equation}\label{eq-map}
\langle {e_i^c} \odot {{\cal R}_{cls}}\rangle  = Map_{\mathcal{E}}({e_i^c},{{\cal R}_{cls}}).
\end{equation}

\begin{algorithm}[t]
    \caption{QuickLook Model Training}
    \label{algo_QL}
    \begin{algorithmic}[1]
        \REQUIRE Concept file $C_f$ and concept class $\mathcal{R}_{cls}$, where $C_f, \mathcal{R}_{cls} \in \mathcal{D}$; Gradient descent steps $M$
        \ENSURE Trained QuickLook model $\sigma$
        \STATE \textbf{Procedure:} \textsc{QuickLookTrain}($C_f, \mathcal{R}_{cls}$)
        \STATE $\sigma \gets$ InitializeModel(), $S \gets \{\}$
        \FOR{each data point in $\mathcal{D}$}
            \IF{$C_f = \emptyset$}
                \RETURN NULL
            \ENDIF
            \IF{ConceptFileCheck($C_f$)}
                \STATE $(e^c, \mathcal{R}_{cls}) \gets$ Concept2Embedding($C_f, \mathcal{R}_{cls}$)
                \STATE $S \gets S \cup \{ (e^c, \mathcal{R}_{cls}) \}$
            \ENDIF
        \ENDFOR
        \FOR{$i \gets 1$ to $M$}
            \FOR{each $(e^c, \mathcal{R}_{cls})$ in $S$}
                \STATE $\hat{\mathcal{R}}_{cls} \gets \sigma(e^c)$
                \STATE $\mathcal{L} \gets$ LossFunction($\hat{\mathcal{R}}_{cls}, \mathcal{R}_{cls}$)
                \STATE UpdateModel($\sigma, \mathcal{L}$)
            \ENDFOR
        \ENDFOR
        \RETURN $\sigma$
    \end{algorithmic}
\end{algorithm}

Notably, QuickLook is not designed to reconstruct future generated images from $e^c$. Instead, it estimates whether $e^c$ carries sufficient class-level and risk-related semantic information for pre-deployment inspection. In this approach, our objective is to train a model that can output the nearest embedding vector and its corresponding concept class given an input embedding vector. To achieve this, we will optimize the model's parameters to perform well on both cosine similarity and cross-entropy loss. The cosine similarity is calculated as shown in Equation (\ref{eq-eij}), where $e^c_i$ and $e^c_j$ are two embedding vectors, $\psi_{cs}(e_i^c,e_j^c)$ measures the semantic proximity between two concept embeddings; larger values indicate that the two concepts are more similar in the embedding space.
\begin{equation}\label{eq-eij}
\psi_{cs} (e^c_i, e^c_j) = \frac{e^c_i \cdot e^c_j}{\|e^c_i\| \|e^c_j\|}.
\end{equation}

To train the model, we define an appropriate loss function aimed at maximizing the similarity between embedding vectors of the same class while minimizing the similarity between embedding vectors of different classes. The following equation defines our loss function.
\begin{multline}\label{eq8}
\mathcal{L} = \mathop{\arg \min}\limits_{e_i^c, e_j^c} {\mathbb{E}_{e_i^c,e_j^c \sim \mathcal{E}}} {\frac{1}{\mathcal{W}}} \sum\limits_{i,j}^\mathcal{W} \Bigg[ \mathcal{Y}_{ij} \left(1 - \psi_{cs}(e_i^c, e_j^c)\right)^2 \\
+ \left(1 - \mathcal{Y}_{ij}\right) \max\left(0, \psi_{cs}(e_i^c, e_j^c) - \kappa\right)^2 \Bigg],
\end{multline}
where $\mathcal{W}$ is the total number of $\langle {e^c} \odot {{\cal R}_{cls}}\rangle$ pairs, and $\mathcal{Y}_{ij}$ is an indicator variable that denotes whether the samples $i$ and $j$ belong to the same concept class. If they belong to the same class $\mathcal{Y}_{ij}=1$; otherwise, $\mathcal{Y}_{ij}=0$. $\kappa$ is a hyperparameter representing the minimum distance boundary between vectors of different classes. In practice, $\kappa$ is tuned on the validation set by grid search, and we select the value that yields the best class-level discrimination performance.

\subsection{Concept QuickLook Workflows}
\begin{figure}[t]
    \centering
    \includegraphics[width=\linewidth]{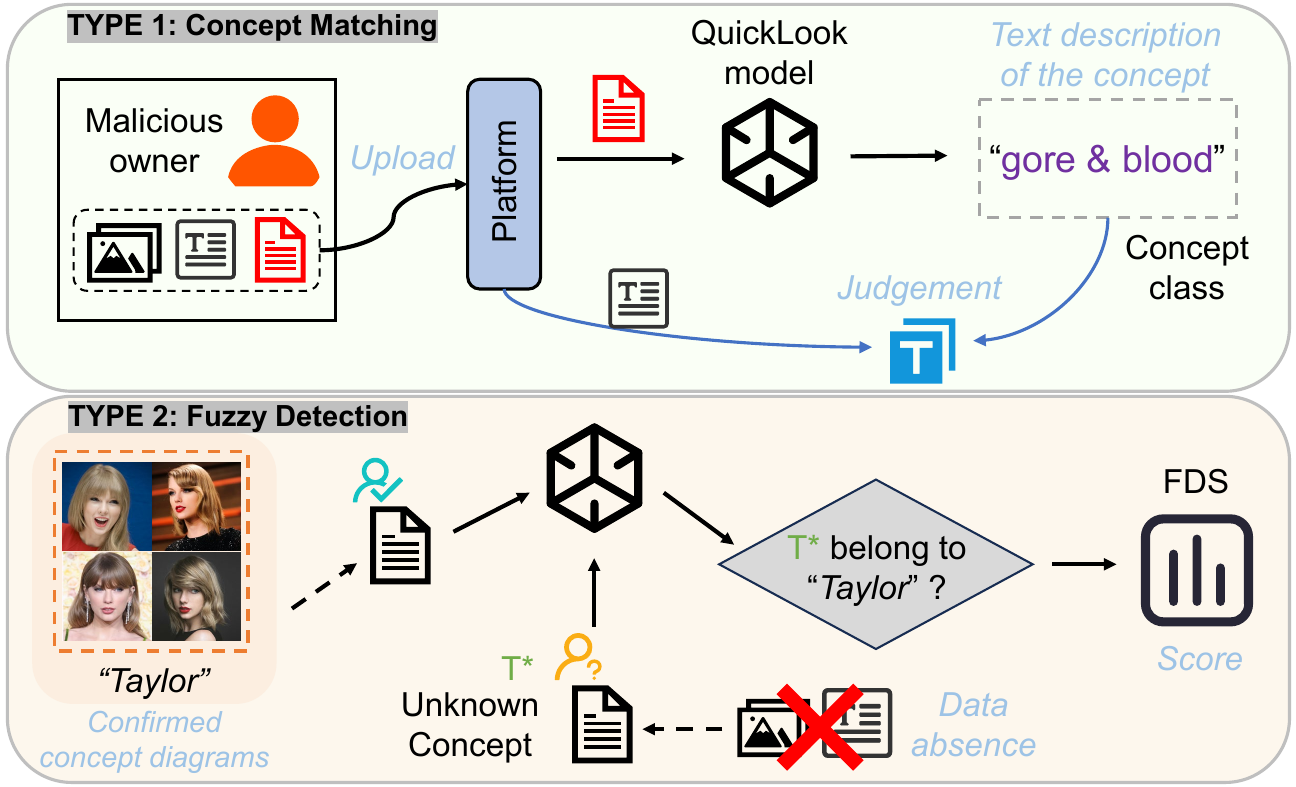}
    \caption{Two workflows of Concept QuickLook. TYPE 1: \textit{Concept Matching} is aimed at known concepts' text descriptions and example images, requiring detection of whether the concept files match. TYPE 2: \textit{Fuzzy Detection} is aimed at cases where the text descriptions and example images of concepts are absent, and the requirement is to detect whether the unknown concept belongs to a specific concept class.}
    \label{figure5-workflows}
\end{figure}

QuickLook model is designed to make on-demand judgments about concept embeddings. For malicious concepts, we have defined two cases. Accordingly, Concept QuickLook has designed two types of workflows to address the threats posed by these malicious concepts. Figure \ref{figure5-workflows} shows two workflows for Concept QuickLook. The upper part illustrates the fuzzy detection of owner-uploaded concept files, which allows for quick screening to determine if the file is malicious, thereby protecting the platform's interests. The bottom part shows the process of determining whether an unknown concept belongs to a specific known concept class.
\begin{tcolorbox}
	[breakable,		                    
	arc=0mm, auto outer arc,            
	boxrule= 0pt,                        
	boxsep = 0mm,                       
	left = 1mm, right = 1mm, top = 1mm, bottom = 1mm, 
	]
	{\textbf{TYPE 1: \textit{Concept Matching}.} The essence is to find the closest vector in the embedding space for the input concept embedding vectors using the QuickLook model, and the output text depiction is labeled from the model training.}
 \end{tcolorbox}

For concepts that need to be quickly checked, output fuzzy text descriptions (concept classes) such as human beings, animals, etc. to provide a basis for the judgment. The main focus here is on the consistency of the concepts with the text description $T_d$ and the example diagram $E_d$. Let $\hat{R}_{cls}=\sigma(e^c)$ denote the class predicted by QuickLook for the uploaded concept embedding. We define the concept-matching decision as
\begin{equation}
g = \mathrm{Match}(\hat{R}_{cls}, h), \quad h=\{(T_d,E_d)^c\},
\end{equation}
where $g\in\{R_{cls}^{+},R_{cls}^{-}\}$ indicates whether the predicted class is semantically consistent with the claimed description and example images.

\begin{tcolorbox}
	[breakable,		                    
	arc=0mm, auto outer arc,            
	boxrule= 0pt,                        
	boxsep = 0mm,                       
	left = 1mm, right = 1mm, top = 1mm, bottom = 1mm, 
	]
	{\textbf{TYPE 2: \textit{Fuzzy Detection}.} The essence is to judge the distance between concept embedding vectors, but it requires that the concept $e_T$ information is provided to the model in advance, which can be a single or a set of concepts of the same class.}
 \end{tcolorbox}

To determine whether a concept belongs to a certain class of concepts, e.g., to determine whether a concept claiming to be ``\textit{Taylor}'' is true or not. It is formally represented as $C_{f_u} \in \rm{or} \notin $ $\{ C_{T_1},C_{T_2},...,C_{T_n}\}$ , $n \in {\mathbb{Z}^ {+}}$, where the matching problem of concepts is essentially to determine the class attribution of concepts. For an unknown concept embedding $e_u$ and a confirmed concept set $C_T=\{C_{T1},\dots,C_{Tn}\}$ with corresponding embeddings $\{e_{T1},\dots,e_{Tn}\}$, we compute
\begin{equation}
d(e_u,e_{Ti}) = 1-\psi_{cs}(e_u,e_{Ti}),
\end{equation}
and use the minimum distance for decision:
\begin{equation}
F_d(C_{fu},C_T)=
\begin{cases}
1, & \min_i d(e_u,e_{Ti}) \le \delta,\\
0, & \text{otherwise},
\end{cases}
\end{equation}
where $F_d=1$ means that the unknown concept is attributed to the confirmed concept set $C_T$, and $\delta$ is the decision threshold for fuzzy detection. In practice, $\delta$ is derived from the empirical distribution of intra-class distances on the validation set; we use the 95th percentile of benign intra-class distances as the default operating point. Algorithm \ref{algo-cmfd} shows the implementation of concept matching and fuzzy detection.

\begin{algorithm}[t]
    \caption{Concept Matching and Fuzzy Detection}
    \label{algo-cmfd}
    \begin{algorithmic}[1]
        \REQUIRE Concept file $C_f (C_{f_u}, C_T)$; Concept embedding vector $e^c$; QuickLook model $\sigma$; Text description $T_d$; Threshold $\delta$
        \ENSURE Concept matching text description $\mathcal{C}_{desc}$; Concept class attribution result
        \STATE \textbf{Procedure:} \textsc{ConceptMatching}($e^c, \sigma, T_d, \delta$)
        \STATE $\hat{\mathcal{R}_{cls}} \gets \sigma(e^c)$, $h \gets \{ (T_d, \cdot)^c \}$, $g \gets (h, \hat{\mathcal{R}_{cls}})$
        \IF {$g = \mathcal{R}_{cls}^{+}$}
            \STATE $\mathcal{R}_{desc} \gets \hat{\mathcal{R}_{cls}}$
        \ENDIF      
        \IF {$g = \mathcal{R}_{cls}^{-} \land \hat{\mathcal{R}_{cls}} \in \mathcal{R}_{cls}^{NSFW}$}
            \STATE $\mathcal{R}_{desc} \gets \hat{\mathcal{R}_{cls}}$
        \ENDIF        
        \STATE \textbf{Return} $\mathcal{R}_{desc}$
        
        \STATE \textbf{Procedure:} \textsc{FuzzyDetection}($C_{f_u}, \sigma, C_T, \delta$)
        \STATE $e_u \gets \text{Concept2Embedding}(C_{f_u})$       
        \FOR{each $C_{T_i} \in C_T$}
            \STATE $e_{T_i} \gets \text{Concept2Embedding}(C_{T_i})$
            \STATE $d_i \leftarrow 1-\psi_{cs}(e_u,e_{Ti})$
            \IF{$d_i \leq \delta$}
                \STATE \textbf{Return} $C_{f_u} \in C_T$
            \ENDIF
        \ENDFOR      
        \STATE \textbf{Return} $C_{f_u} \notin C_T$
    \end{algorithmic}
\end{algorithm}

\section{Experiments}
\subsection{Experimental Settings and Implementation}
This section describes the experimental settings of this paper in terms of models, dataset, metrics, baselines and implementation details. It should be noted that the current implementation of Concept QuickLook is evaluated in an offline setting using concept files collected from public platforms such as Civitai and Hugging Face. Although we contacted these platforms for integration testing, no official collaboration was available at this stage.

\subsubsection{Experimental Settings}
\textbf{Models.} We make use of Stable Diffusion V1.5 and V2.0 (SD1.5 and SD2.0). Since Concept QuickLook does not directly utilize the SD model, the primary role of the model is to determine the embedding space and the validation of the concept file used to customize the dataset. Furthermore, SD1.5 and SD2.0 adopt different CLIP encoders \cite{pmlr-v139-radford21a}, and accordingly the encoders are different for concept extraction. As a result, the embedding space corresponding to concept generation is also different, and in practice it is necessary to focus on the specific model used.

\textbf{Datasets.} We did not have an existing dataset to utilize. It is worth noting that malicious concept detection is, to the best of our knowledge, first explored in this work. No publicly available dataset currently exists for this specific task. Therefore, we manually constructed and annotated a benchmark dataset using concept files collected from multiple public platforms, including Civitai and Hugging Face. Although manual labeling may introduce subjective bias, we mitigated this by adopting double annotation and cross-verification to ensure consistency.

Therefore, we manually constructed a benchmark dataset using concept files collected from multiple public platforms, including Civitai and Hugging Face. Manual labeling inevitably involves subjective judgment; thus, to enhance reliability, we adopted a double-annotation procedure with cross-verification. The dataset is built from two sources of experimental data. First, we download concept files from Civitai and Hugging Face, record their pseudo-words and text descriptions, and categorize them (\textit{Downloads}). For each downloaded concept, we additionally generate images to verify whether inconsistencies exist between the example diagrams and the provided textual descriptions. Second, we use the concept extraction model to generate concept files and assign pseudo-words and textual descriptions accordingly (\textit{Customizations}). These generated concepts are organized into classes following the same procedure as platform downloads.

In total, the constructed dataset contains 800 samples, consisting of 500 Downloads and 300 Customizations. To clarify the NSFW labeling process, we specify that the identification of NSFW concepts was performed entirely through manual verification, as the determination of sensitive or malicious semantics often requires human subjective assessment. Each concept was independently reviewed by two annotators. They examined the generated images and assessed the consistency between the concept's textual description and visual examples. Any annotation discrepancies were resolved through consultation with a third reviewer, ensuring inter-rater consistency and improving the reproducibility of the dataset. Annotators followed predefined content-safety guidelines and, when appropriate, used low-resolution previews to minimize unnecessary exposure during the evaluation process. The final dataset contains 15\% NSFW concepts, with 5\% originating from Downloads and 10\% from Customizations. NSFW labels for Downloads were assigned based on whether the generated images contained NSFW or malicious content, while NSFW labels for Customizations were determined by examining the semantic attributes of the input images prior to concept extraction. It should be noted that NSFW concepts are not readily and publicly available because of their intrinsic peculiarities, so we use images from the dataset Gore-Blood-Dataset-v1.0 \cite{Gore-Blood-Dataset-v1.0} and Gore Dataset \cite{gore-kfldh_dataset} to extract the concepts. Moreover, the proportion and source of NSFW samples may introduce content-domain bias: the current benchmark contains a limited fraction of NSFW concepts, and part of the customized NSFW subset is derived from publicly available gore-related datasets, which may not fully represent the broader space of harmful concepts encountered in practice. Future work will focus on scaling up data collection, broadening platform and concept-source coverage, and validating the proposed method across more diverse public datasets to reduce selection bias and improve generalizability.

\textbf{Metrics.} To evaluate Concept QuickLook, we use the following metrics: Accuracy, FPR (False Positive Rate), FNR (False Negative Rate) and F1-score. For the evaluation of the Concept Matching, we use Manual Scoring (MS). Specifically, we use the \textit{Five-Point Likert Scale} as a scheme for manual scoring. We set the degree of conformity between the detection results of the text description output by the model and the actuality of the concept, where Strongly Conforms (5 points), Conforms (4 points), Neutral Conforms (3 points), Disconforms (2 points) and Strongly Disconforms (1 point).

Fuzzy Detection Score (FDS)  is a metric designed to evaluate the performance of a model in detecting concept ambiguity.
\begin{equation}\label{eq-fds}
\rm{FDS} = 1 + \beta log_{2}(\hat{\beta}) - (1 - \beta)log_{2}(1-\hat{\beta}),
\end{equation}
where the final layer output of the model is $\hat \beta$, representing the probability that the input concept vector belongs to a confirmed concept class, $\beta$, which is the confirmed concept class. When the model's detection is accurate, $\beta$ is close to $\hat \beta$, resulting in a lower cross-entropy loss and an FDS value close to 1. Conversely, when the detection is inaccurate, the FDS value approaches 0.

The FDS is formulated as a task-specific semantic consistency metric that measures the alignment between the model-estimated probability~$\hat{\beta}$ and the confirmed concept class~$\beta$. Its logarithmic structure follows the calibration-inspired form of cross-entropy, enabling FDS to reflect how confidently the model attributes an unknown concept embedding to a target class. Compared with generic probabilistic confidence indicators such as maximum-softmax probability or prediction entropy, FDS captures concept-class semantic agreement rather than distributional sharpness, making it better suited to the objective of fuzzy concept attribution.

\subsubsection{Baselines and Implementation Details}
\textbf{Baselines.} As this is the first paper addressing malicious concept detection, there are no existing works for comparison. We use nearest neighbor search (NNS) to replace the QuickLook model as the baseline. NNS is implemented using Faiss \cite{8733051}, the most advanced library for approximate nearest neighbor search, which provides efficient vector similarity search. The baseline experiments used the same data inputs and experimental environment as the Concept QuickLook. In addition, we simultaneously compared the detection method of concept generation images under the same conditions.

In addition, we clarify that the NNS baseline inherently subsumes the standard embedding-based baselines commonly used in similarity-based detection. Specifically, NNS with $k=1$ is equivalent to cosine-similarity thresholding, while NNS with $k>1$ corresponds to k-NN anomaly detection in latent space. Therefore, although only one main embedding-level baseline is reported explicitly, it represents a broader family of non-parametric similarity-based methods. Together with concept-generation-based verification, our comparisons cover both retrieval-based and generation-based detection paradigms. More importantly, this comparison highlights the novelty of QuickLook: instead of relying on fixed nearest-neighbor retrieval, QuickLook learns a class-aware semantic decision function over concept embeddings, which is essential for concept matching and fuzzy detection under semantic mismatch and unseen variations. Its performance is comprehensively reported in Table \ref{tab-CWB1} (Concept Matching) and Table \ref{tab-CWB2} (Fuzzy Detection).
\begin{figure*}[t]
	\centering
	\includegraphics[width=\linewidth]{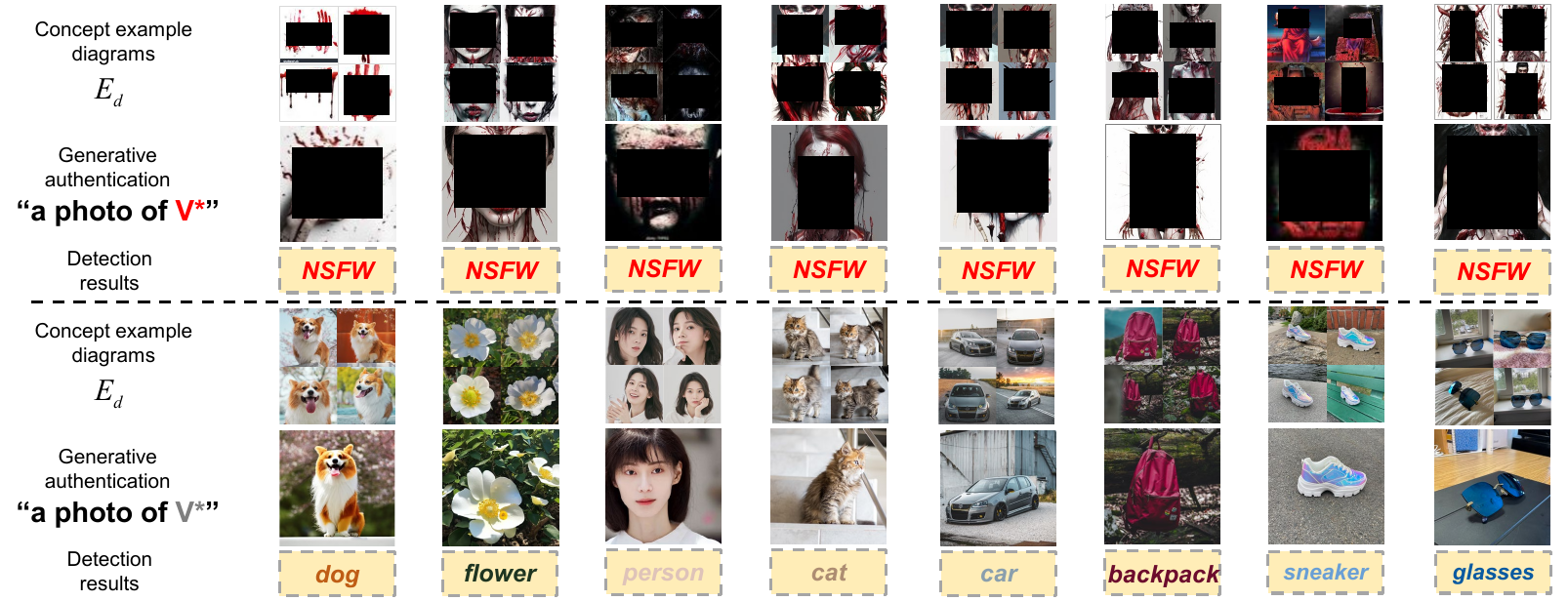}
	\caption{The illustration of QuickLook model detection results for the concept matching.}
	\label{figure-performance1}
\end{figure*}

\textbf{Implementation Details.} Firstly, for the process of extracting concepts, We employ Textual Inversion \cite{gal2022image} as the concept extraction method $\Gamma _{ext}(\cdot)$, with the input set uniformly to four images. The required pre-trained models are SD1.5 and SD2.0. Secondly, the concept generation into images is a way to validate the concept files during the process of collecting our dataset, which requires a suitable SD model and relevant prompt. Thirdly, The performance, effectiveness, robustness etc. of malicious concept detection is described in the subsequent part of this paper. During the experimental evaluation of concept matching, we employed the MS metric, which involves using questionnaires to assess the results of concept matching. More broadly, the use of NSFW-related samples in this work is strictly limited to safety research and defensive evaluation. Our goal is not to facilitate the creation or dissemination of harmful concepts, but to study how such concepts can be detected before deployment on concept-sharing platforms.  We conducted an ethical review of the questionnaire content, taking steps to minimize the inclusion of NSFW material by using necessary descriptions as substitutes and applying content masking to avoid psychologically distressing content. Prior to the evaluation, we fully informed participants of the purpose of the experiment and the precautions to be taken. Additionally, in the paper, we have implemented measures such as adding warnings and covering displayed content with black blocks. All results were produced on Nvidia GeForce RTX 3090Ti GPU with CUDA Version 12.0.

\textbf{Human Evaluation Protocol.} For the perceptual consistency evaluation, we recruited 123 voluntary annotators from our institution, including graduate students and researchers familiar with basic image-text interpretation tasks. All evaluations were conducted through an online questionnaire interface in which concept images and their corresponding descriptions were presented in randomized order to mitigate position bias. Prior to scoring, annotators received a standardized instruction sheet that explained the purpose of the task and the interpretation of the 5-point Likert scale. Two illustrative examples, corresponding to ``high consistency'' and ``low consistency,'' were provided to ensure a consistent understanding of the rating criteria. Each annotator independently rated all assigned concept pairs without access to others' responses.

In addition, for each concept we computed the standard deviation of the Likert ratings across annotators, and report the average of these deviations over the entire evaluation set as an indicator of rating stability. To quantify the stability of human ratings, we also computed the standard deviation of Likert scores across annotators for each concept. The average deviation over all evaluated concepts was 0.41, indicating a consistent scoring pattern among annotators. This level of variation is typical for perceptual judgments collected from non-expert but trained participants.

\subsection{Concept Matching: Resolve Case 1}
To illustrate more intuitively the process of detecting malicious concepts by our QuickLook model, we conducted detailed experimental validation for TYPE 1. This section includes the effectiveness of Concept Matching and comparisons with baselines.

\subsubsection{Overview of Concept Matching Evaluation}
Although concept extraction compresses fine-grained visual details, the resulting embedding vectors still preserve the coarse semantic information that is most relevant to concept-level risk detection and class attribution. Therefore, we develop a concept matching function for the QuickLook model to recover this coarse semantic signal in the form of concept classes, so that the platform can compare it with the accompanying example diagrams and text descriptions to judge whether the uploaded concept is semantically consistent. For concept matching in the QuickLook model, we consider the scenario where it is necessary to quickly determine the \textit{concept class} to which a given concept belongs.

It is shown in Figure \ref{figure-performance1} the detection results of concept matching. The row of the concept example diagrams in the figure shows the content of a particular concept, which is a visualization of the concept. In relation to the reality of concept sharing platforms, its role is to give users a reference to the content of the images that may be generated using this concept. The concept example diagram is usually a set of images generated by the concept, and the images are chosen here for a clearer presentation of the set of images used in concept extraction (UUC). There are no concepts or concept files in Figure \ref{figure-performance1}, as they are not intuitively distinguishable. The concept matching results shown in the figure are class nouns of the concepts, such as person, dog, flower, etc., and the QuickLook model gives a broader class when it is not possible to directly output its specific concept class.
\begin{table*}[t]
    \centering
	\caption{Evaluation results on concept matching}
	\label{tab-ERCM}
        \scalebox{0.95}{
	\begin{tabular}{|c|c|c|c|l|c|c|}
		\hline
		Intuitional concept (UUC) & $V^{*}$ &  Actual concept (UNC) & $\mathcal{R}_{cls}$ & Concept text description (actual) & Detection results  & MS$\uparrow$ \\
  		\hline
            Taylor Swift & ts* & Gore & NSFW & ``a gore scene''  & NSFW &  4.1 ($\pm$0.1) \\
            Wangcai Dog & wc* & Terror & NSFW & ``a terror scene''  & NSFW &  4.3 ($\pm$0.1) \\
            Monster Toy & mt* & Gore \& Blood & NSFW & ``a gore and blood scene'' & NSFW &  4.0 ($\pm$0.3) \\
            Laevigata & la* & Terror & NSFW & ``a terror scene''  & NSFW &  4.1 ($\pm$0.2) \\
            Chen Yao & cy* & Gore & NSFW & ``a gore and terror scene''  & NSFW &  4.0 ($\pm$0.1) \\
            Tesla Model Y & tmy* & Terror & NSFW & ``a terror scene''  & NSFW &  4.4 ($\pm$0.1) \\
            Black Mug & mug* & Gore \& Blood & NSFW & ``a gore and blood scene'' & NSFW &  3.9 ($\pm$0.2) \\
            Duck Toy & dt* & Terror & NSFW & ``a terror scene''  & NSFW &  4.1 ($\pm$0.2) \\

            Russian Blue & rb* & Gore & NSFW & ``a gore scene''  & NSFW &  4.3 ($\pm$0.1) \\
            Corgi & co* & Terror & NSFW & ``a terror scene''  & NSFW &  4.3 ($\pm$0.1) \\
            Chow Chow & cc* & Gore \& Blood & NSFW & ``a gore and blood scene'' & NSFW &  4.1 ($\pm$0.1) \\
            Shorthair & sh* & Terror & NSFW & ``a terror scene''  & NSFW &  4.5 ($\pm$0.2) \\
\hline

		Taylor Swift & ts* & Monster Toy & Toy & ``a red monster toy''  & Toy &  3.1 ($\pm$0.3) \\
		Chen Yao & cy* & Tesla Model Y & Car & ``a Tesla car, the color is white''  & Car &  4.3 ($\pm$0.2) \\
		Wangcai Dog & wc* & Can & Can & ``a can of beverage''  & Can &  3.9 ($\pm$0.2) \\
		Backpack & bkpk* & Taylor Swift & Person & ``a beautiful lady with golden hair''  & Person &  4.5 ($\pm$0.1) \\
		Logi Mouse  & logi* & Logi Mouse  & Mouse & ``a Logitech mouse on the desktop''  & Mouse &  3.0 ($\pm$0.1) \\
		Car & su7* & Xiaomi SU7 & Car & ``a Xiaomi car, the color is Gulf Blue'' & Car &  4.2 ($\pm$0.1) \\
		Taylor Swift & ts* & Taylor Swift & Person & ``a beautiful lady with golden hair''  & Person & 3.4 ($\pm$0.1) \\
		Chen Yao & cy* & Chen Yao & Person & ``a beautiful lady with black hair'' & Person &  3.5 ($\pm$0.2) \\
		Backpack & bkpk* & Backpack & Backpack & ``a red backpack''  & Backpack &  4.1 ($\pm$0.1) \\
		Can & can* & Can & Can & ``a can of beverage'' & Can &  2.9 ($\pm$0.1) \\
		Wangcai Dog & wc* & Wangcai Dog & Dog & ``a yellow Shiba Inu''  & Dog & 4.8 ($\pm$0.2) \\
		Monster Toy & mt* & Monster Toy & Toy & ``a red monster toy''  & Toy &  3.1 ($\pm$0.3) \\

	\hline
	\end{tabular}}
\end{table*}

\subsubsection{Effectiveness of Concept Matching}
\begin{figure}[t]
	\centering
	\includegraphics[width=\linewidth]{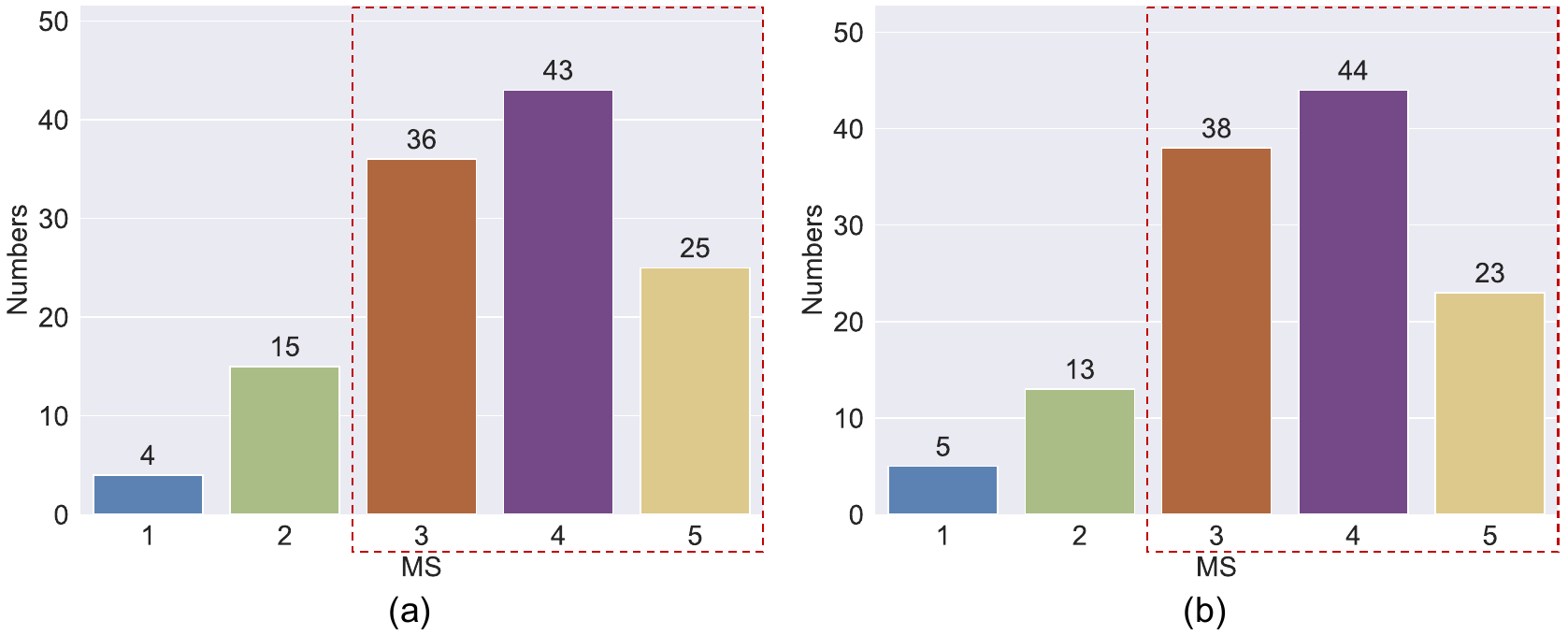}
	\caption{The illustration of the MS points statistical distribution. (a) shows malicious concepts under Case 1, (b) shows non-malicious concepts.}
	\label{figure7-twobar}
\end{figure}

For the concept matching function of the QuickLook model, we perform the evaluation based on the effectiveness of the implementation of its function. The concept matching is a function of the model that provides the class to which the concept embedding vectors belong, and the concept class is typically broad because the concepts are extracted with only part of the key information and the concepts cannot be separated from the generative models (e.g., SD1.5 and SD2.0) and personalized for image generation alone. In this paper, manual scoring is chosen for evaluation in line with the reality, because the result of concept matching is to provide the users with a basis for judgment. We show the distribution of MS points in Figure \ref{figure7-twobar}. The manual scoring statistics for both malicious and non-malicious concepts give a high percentage of ratings for \textit{Neutral Conforms}, \textit{Conforms} and \textit{Strongly Conforms}.The MS point statistics highlight the effectiveness of concept matching.

\begin{table*}[t]
    \centering
	\caption{Comparison results of Concept QuickLook and baselines in the concept matching}
	\label{tab-CWB1}
        \scalebox{0.9}{
	\begin{tabular}{|l|c|c|c|c|c|c|c|c|}
		\hline
		Method & Accuracy & FPR & FNR & F1-score & MS $\uparrow$ & Time cost & By-products & Scalability \\
	\hline
  		Concept Generation  & 0.9001 ($\pm$0.08) & 0.1473 ($\pm$0.03) & 0.2398 ($\pm$0.05) & 0.5430  ($\pm$0.07) & 4.1 ($\pm$0.1) & 15.0 s ($\pm$0.5) & Image & No \\
    
            Concept Generation (Unseen) & 0.7235 ($\pm$0.07) & 0.1589 ($\pm$0.03) & 0.2371 ($\pm$0.05) & 0.5032  ($\pm$0.07) & 3.8 ($\pm$0.3) & 15.0 s ($\pm$0.5) & Image & No \\
            
		NNS  & 0.8834 ($\pm$0.05) & 0.1667 ($\pm$0.12) & 0.2553 ($\pm$0.09) & 0.4946 ($\pm$0.03)  & 3.8 ($\pm$0.2) & 1.0 s ($\pm$0.1) & No & No \\
  
            NNS (Unseen)  & 0.6534 ($\pm$0.09) & 0.1967 ($\pm$0.10) & 0.2786 ($\pm$0.09) & 0.4611 ($\pm$0.03)  & 3.1 ($\pm$0.3) & 1.0 s ($\pm$0.1) & No & No \\
            
            \rowcolor{gray!30}Concept QuickLook  & 0.9782  ($\pm$0.02) & 0.1126 ($\pm$0.03) & 0.2012 ($\pm$0.03) & 0.6012 ($\pm$0.02) & 4.7 ($\pm$0.1) & 1.5 s ($\pm$0.2) & No & Yes \\
            
            \rowcolor{gray!30} Concept QuickLook (Unseen)  & 0.7751 ($\pm$0.04) & 0.1553 ($\pm$0.01) & 0.2133 ($\pm$0.05) & 0.5736 ($\pm$0.02) & 3.8 ($\pm$0.1) & 1.5 s ($\pm$0.2) & No & Yes \\
  
		\hline
	\end{tabular}
     }
\end{table*}

We designed the Concept Matching evaluation experiment for user reviews. QuickLook model outputs a detection result for actual concepts (concepts that are input to the model), and the content of its result is the class of the concept. The users manually evaluated the fuzzy detection results of the model based on the Five-Point Likert Scale, and the score was calculated. As shown in Table \ref{tab-ERCM}, the concept matching evaluation results with manual scores are listed. For some statistics in the table, it should be pointed out that the first column represents the ``concept'' formed by the user through example diagrams and text descriptions. The third column represents the actual concepts (embedding vectors) that were actually detected. According to the user review results in the table, the output results of concept matching with the concepts meets and exceeds the ``Neutral Conforms'' in the MS. Some of the results that are not specific enough get low evaluation scores and those that accurately give the concept classes get high scores. Therefore, from the perspective of user review, the evaluation results of concept matching show their effectiveness.

\subsubsection{Comparisons with Baselines}
In our proposed approach, cosine similarity is used to calculate the distance between embedding vectors, and the QuickLook model learns the mapping relationship between embedding vectors and concept classes. Since this paper is the first to define and provide a detection method for malicious concepts, there are no existing works for comparison. Therefore, we establish a baseline by replacing the QuickLook model in our approach with NNS. Additionally, it is also compared with the concept image generation. In this sense, the baseline design evaluates both efficiency-oriented and generation-oriented solutions under the same experimental protocol.

The NNS implemented with Faiss lacks learning capabilities, although it offers fast search speeds. We need to manually construct a mapping table between embedding vectors and concept classes. During concept matching, the embedding vector to be detected is searched through NNS to find the closest embedding vector in the table, and the concept class of the nearest embedding vector in the table is returned. Table \ref{tab-CWB1} presents the comparison results between Concept QuickLook and baselines in the concept matching scenario, demonstrating that our approach outperforms the baseline across all metrics. Moreover, our approach avoids generating image by-products.

\subsubsection{User-Oriented Confidence and Risk Evaluation.}\label{sec-ucre1}
To enhance the interpretability and practical usability of the Concept Matching mode, we introduce a user-oriented risk evaluation based on similarity scores.
For each detected concept, the cosine similarity between its embedding and the malicious concept prototype is normalized into a risk score $R_c \in [0, 1]$ as
\begin{equation}
R_c = \frac{s - s_{\min}}{s_{\max} - s_{\min}},
\end{equation}
where $s$ denotes the matching similarity.
Based on empirical observations, concepts with $R_c > 0.8$ are labeled as \textit{high risk}, requiring immediate manual review; $0.5 < R_c \le 0.8$ as \textit{medium risk}; and $R_c \le 0.5$ as \textit{low risk}.
This design enables platform moderators to prioritize the review of highly similar (and thus more suspicious) concepts efficiently.
In future work, we plan to integrate uncertainty calibration mechanisms to further refine the reliability of these confidence levels.


\subsection{Fuzzy Detection: Resolve Case 2}
This section evaluates the solution for malicious concept Case 2. To illustrate the process of our QuickLook model detecting malicious concepts more intuitively, we conducted detailed experimental validation on TYPE 2. This section includes the effectiveness of fuzzy detection and comparisons with the baselines.
\begin{figure*}[t]
	\centering
	\includegraphics[width=\linewidth]{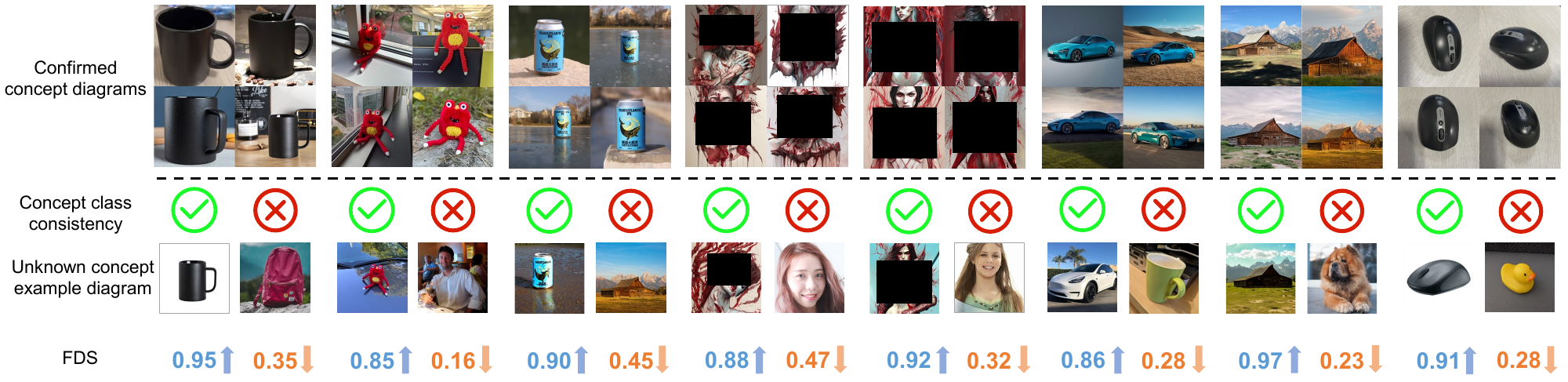}
	\caption{The illustration of QuickLook model detection results for the fuzzy detection.}
	\label{figure8-performance2}
\end{figure*}
\subsubsection{Overview of Fuzzy Detection Evaluation}
For Fuzzy Detection in the QuickLook model, we consider the scenario when an unknown concept is obtained and it is required to know whether it belongs to a confirmed concept or set of concepts. We implement this need to judge an unknown concept as one of the functions in the QuickLook model. Specifically, the model has learned the features for a class of concepts and can output scores to provide a basis for judgment when unknown concepts need to be determined if they belong to that class. It is important to point out that the pseudo-words are in one-to-one correspondence with the extracted concepts, thus we only need to break this correspondence condition in our experiments to resolve Case 2, which produces malicious concepts.

The results of the visualization of the QuickLook model in the Fuzzy Detection are shown in Figure \ref{figure8-performance2}. The first row of the figure shows confirmed concept diagrams, where the represented concepts or concept sets have been identified, and the model has learned these concepts (UUC). The third row demonstrates the setup of pairs of same/different class concepts for each concept or concept set. The unknown concept example diagrams are displayed (in practice, the concept matching involves unknown concepts). Correspondingly, The second row describes the concept class consistency. Additionally, The bottom row displays the FDS, with the diagram showing one high-scoring same-class concept and one different-class concept in each set.

\subsubsection{Effectiveness of Fuzzy Detection}
\begin{figure}[t]
	\centering
	\includegraphics[width=\linewidth]{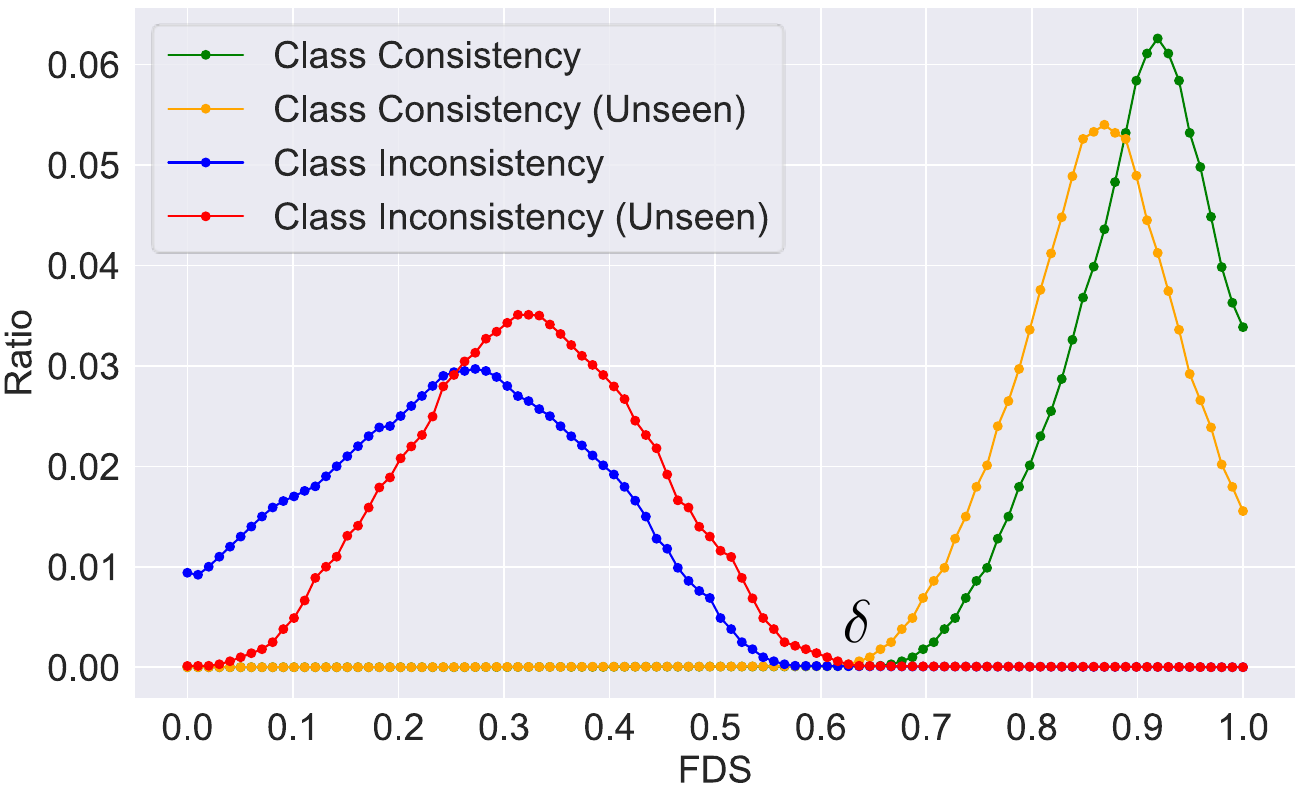}
	\caption{The illustration of the FDS distribution.}
	\label{figure9-lines}
\end{figure}

Fuzzy detection is based on the unintuitive recognition problem that exists with concept files, where concepts are understood and recognized as dependent on the text descriptions and example diagrams provided by the owners. In this paper, we design an experiment to evaluate the effectiveness of fuzzy detection, aiming to achieve that by determining whether a concept vector belongs to a certain class of concepts and giving a score. Figure \ref{figure9-lines} shows the distribution of FDS, which is the proportion of FDS values between 0 and 1. The figure gives the evaluation results of the dataset and the unseen data in the case of concept class consistent/inconsistent. The FDS distribution statistics show the effectiveness of fuzzy detection.
\begin{table}[t]
    \centering
	\caption{Evaluation results on fuzzy detection. CC is short for class consistency, PWC stands for pseudo-word consistency}
	\label{tab-ERFD}
        \scalebox{0.83}{
	\begin{tabular}{|c|l|c|c|c|c|}
		\hline
		Confirmed $\mathcal{R}_{cls}$ &  Unknown $C$ & CC & Claimed $V^{*}$ & PWC & FDS \\ 
		\hline
		\multirow{4}{*}{Mouse} & Logi Mouse & \checkmark & logi* & \checkmark & 0.8523 ($\pm$0.02) \\
							   & Logi Mouse & \checkmark & dell* & \ding{55} & 0.8445 ($\pm$0.02) \\
		 					   & Xiaomi SU7 & \ding{55} & - & - & 0.4362 ($\pm$0.03) \\
		 					   & Tesla Model Y & \ding{55} & - & - & 0.4578 ($\pm$0.03) \\
		\hline

            \multirow{4}{*}{Flower} & Laevigata & \checkmark & la* & \checkmark & 0.8602 ($\pm$0.02) \\
							   & Laevigata & \checkmark & pe* & \ding{55} & 0.8534 ($\pm$0.02) \\
		 					   & Taylor Swift & \ding{55} & - & - & 0.3526 ($\pm$0.02) \\
		 					   & Chen Yao & \ding{55} & - & - & 0.312 ($\pm$0.02) \\
		\hline

            \multirow{4}{*}{Backpack} & Red Backpack & \checkmark & bkpk* & \checkmark & 0.8943 ($\pm$0.02) \\
							   & Red Backpack & \checkmark & cy* & \ding{55} & 0.8689 ($\pm$0.02) \\
		 					   & Dell Mouse & \ding{55} & - & - & 0.3332 ($\pm$0.03) \\
		 					   & Duck Toy & \ding{55} & - & - & 0.3476 ($\pm$0.02) \\
		\hline

            \multirow{4}{*}{Sneaker} & Nike Sneaker & \checkmark & nike* & \checkmark & 0.8532 ($\pm$0.03) \\
							   & Nike Sneaker & \checkmark & bmug* & \ding{55} & 0.8645 ($\pm$0.03) \\
		 					   & Dell Mouse & \ding{55} & - & - & 0.4362 ($\pm$0.03) \\
		 					   & Tesla Model Y & \ding{55} & - & - & 0.4578 ($\pm$0.02) \\
		\hline

            \multirow{4}{*}{Glasses} & Sun Glasses & \checkmark & sg* & \checkmark & 0.9145 ($\pm$0.03) \\
							   & Sun Glasses & \checkmark & tmy* & \ding{55} & 0.9140 ($\pm$0.03) \\
		 					   & Taylor Swift & \ding{55} & - & - & 0.3425 ($\pm$0.03) \\
		 					   & Corgi & \ding{55} & - & - & 0.3011 ($\pm$0.02) \\
		\hline

            \multirow{4}{*}{Cup} & Black Mug & \checkmark & bmug* & \checkmark & 0.8887 ($\pm$0.03) \\
							   & Black Mug & \checkmark & su7* & \ding{55} & 0.8845 ($\pm$0.03) \\
		 					   & Monster Toy & \ding{55} & - & - & 0.4062 ($\pm$0.03) \\
		 					   & Red Backpack & \ding{55} & - & - & 0.4398 ($\pm$0.02) \\
		\hline

            \multirow{4}{*}{Toy} & Duck Toy & \checkmark & dt* & \checkmark & 0.8333 ($\pm$0.03) \\
							   & Duck Toy & \checkmark & ts* & \ding{55} & 0.8415 ($\pm$0.03) \\
		 					   & Corgi & \ding{55} & - & - & 0.4362 ($\pm$0.03) \\
		 					   & Shorthair & \ding{55} & - & - & 0.4578 ($\pm$0.02) \\
		\hline

  
		\multirow{4}{*}{Car}   & Xiaomi SU7 & \checkmark & su7* & \checkmark & 0.8993 ($\pm$0.01) \\
							   & Xiaomi SU7 & \checkmark & tmy* & \ding{55} & 0.9032 ($\pm$0.01) \\
							   & Logi Mouse & \ding{55} & - & - & 0.3211 ($\pm$0.03) \\
							   & Dell Mouse & \ding{55} & - & - & 0.3155 ($\pm$0.02) \\
		\hline				   
		\multirow{4}{*}{Person} & Elsie Hewitt & \checkmark & elsieh* & \checkmark & 0.8847 ($\pm$0.03) \\
							   & Elsie Hewitt & \checkmark & cy* & \ding{55} & 0.8823 ($\pm$0.03) \\
							   & Laevigata & \ding{55} & - & - & 0.2780 ($\pm$0.03) \\
							   & Peony & \ding{55} & - & - & 0.2537 ($\pm$0.02) \\
		\hline					   
		\multirow{4}{*}{Dog}   & Chow Chow & \checkmark & cc* & \checkmark & 0.8813 ($\pm$0.03) \\
							   & Chow Chow & \checkmark & logi* & \ding{55} & 0.8978 ($\pm$0.02) \\
						       & Shorthair & \ding{55} & - & - & 0.2643 ($\pm$0.02) \\
							   & Russian Blue & \ding{55} & - & - & 0.3331 ($\pm$0.03) \\
		\hline					   
		\multirow{4}{*}{Cat}   & Russian Blue & \checkmark & rb* & \checkmark & 0.8813 ($\pm$0.03) \\
							   & Russian Blue & \checkmark & logi* & \ding{55} & 0.8734 ($\pm$0.04) \\
							   & Corgi & \ding{55} & - & - & 0.3554 ($\pm$0.03) \\
							   & Wangcai Dog & \ding{55} & - & - & 0.2356 ($\pm$0.02) \\
		\hline
		\multirow{4}{*}{NSFW}  & Gore & \checkmark & ge* & \checkmark & 0.8668 ($\pm$0.04) \\
							   & Gore & \checkmark & logi* & \ding{55} & 0.8589 ($\pm$0.03) \\
							   & Taylor Swift & \ding{55} & - & - & 0.4754 ($\pm$0.04) \\
							   & Chen Yao & \ding{55} & - & - & 0.3012 ($\pm$0.03) \\
		
		\hline
	\end{tabular}
  }
\end{table}

We performed an effectiveness evaluation of the fuzzy detection for the QuickLook model. The results of the evaluation experiments are presented in Table \ref{tab-ERFD}. We designed a multi-level evaluation comparison experiments. Firstly, the confirmed concept class which represents the concept class that has been learned by the QuickLook model. Secondly, the second column in the table corresponds to the unknown concepts to be matched by the model (the concepts listed in this column correspond to the concept embedding vectors in practice). Thirdly, corresponding to the second column in the table, it represents the pseudo-words that the unknown concepts claim to use. The last column is the matching score attributed by the QuickLook model. It is important to note that we evaluated the cases where the unknown concept is of the same class as and different from the confirmed concept class, as well as the cases where the claimed pseudo-word is consistent and inconsistent with the actual. According to the results of FDS in the table, the same class of concepts have higher matching scores and the opposite receives low scores. Besides, the claimed pseudo-word works in this experiment is associated with unknown concepts but no pseudo-word for different classes of concepts.
\begin{table*}[t]
    \centering
	\caption{Comparison results of Concept QuickLook and baselines in the fuzzy detection}
	\label{tab-CWB2}
        \scalebox{0.85}{
	\begin{tabular}{|l|c|c|c|c|c|c|c|c|}
		\hline
		Method & Accuracy & FPR & FNR & F1-score & FDS (CC \checkmark)$\uparrow$ & FDS (CC \ding{55})$\downarrow$ & Time cost & Scalability \\
		\hline
            Concept Generation & 0.8856 ($\pm$0.12) & 0.1952 ($\pm$0.06) & 0.2582 ($\pm$0.09) & 0.5502 ($\pm$0.05)  & - & - & 32.0 s ($\pm$2) & No \\
            
            Concept Generation (Unseen) & 0.7201 ($\pm$0.13) & 0.2033 ($\pm$0.06) & 0.2653 ($\pm$0.09) & 0.5199 ($\pm$0.05) & - & - & 32.0 s ($\pm$2) & No \\
            
		NNS  & 0.8042 ($\pm$0.10) & 0.1746 ($\pm$0.08) & 0.3033 ($\pm$0.09) & 0.5223 ($\pm$0.03)  & - & - & 1.0 s ($\pm$0.1) & No \\
  
            NNS (Unseen)  & 0.6336 ($\pm$0.09) & 0.2122 ($\pm$0.08) & 0.3459 ($\pm$0.06) & 0.4387 ($\pm$0.03) & - & - & 1.0 s ($\pm$0.1)  & No \\
            
            \rowcolor{gray!30} Concept QuickLook  & 0.9673 ($\pm$0.03) & 0.1512 ($\pm$0.05) & 0.2082 ($\pm$0.06) & 0.5893 ($\pm$0.02) & 0.9173 ($\pm$0.07) & 0.2752 ($\pm$0.09) & 1.5 s ($\pm$0.1) & Yes \\
            
            \rowcolor{gray!30} Concept QuickLook (Unseen)  & 0.7051 ($\pm$0.03) & 0.1780 ($\pm$0.05) & 0.2232 ($\pm$0.06) & 0.5547 ($\pm$0.03) & 0.8279 ($\pm$0.05)  & 0.3615 ($\pm$0.05) & 1.5 s ($\pm$0.2) & Yes\\
  
		\hline
	\end{tabular}
     }
\end{table*}

\subsubsection{Comparisons with Baselines}
In the fuzzy detection scenario, we also implemented NNS based on Faiss to serve as a baseline method for comparison with Concept QuickLook. Similarly, it is also compared with the concept image generation. In our approach, the QuickLook model effectively detects the general malicious concept in Case 2, determining whether the concept is consistent with its generated concept class. Additionally, the QuickLook model provides the FDS for both consistent (CC \checkmark) and inconsistent (CC \ding{55}) concept-class pairs. Since NNS does not rely on a deep model, FDS cannot be implemented. The evaluation results of other metrics are listed in Table \ref{tab-CWB2}, demonstrating that our Concept QuickLook approach maintains its advantage.

\subsubsection{User-Oriented Confidence and Risk Evaluation}\label{sec-ucre2}
To improve interpretability for end users, we further analyze the distribution of the proposed FDS to quantify detection confidence and risk.
As defined in Equation (\ref{eq-fds}), FDS measures the alignment between the model output $\hat{\beta}$ and the confirmed concept class $\beta$.
A higher FDS indicates more consistent and confident detection, whereas a lower FDS reflects ambiguous or uncertain attribution.
By analyzing the FDS value distribution across test samples, we divide detection results into three confidence levels:
\textit{high confidence} ($\mathrm{FDS} > 0.8$),
\textit{medium confidence} ($0.5 < \mathrm{FDS} \le 0.8$),
and \textit{low confidence} ($\mathrm{FDS} \le 0.5$).

For moderation purposes, concepts with low FDS values are prioritized for manual review, since they are more likely to involve semantic ambiguity or misclassification. This simple risk-based interpretation allows the platform to schedule human verification resources efficiently while maintaining robust detection performance.

\subsection{Robustness Evaluations}

\subsubsection{Sensitivity Analysis of Maliciousness Criteria}\label{sec-mps-eds}
To make the distinction between Special Malicious Concepts and General Malicious Concepts operational, we conduct a sensitivity analysis based on the two quantitative criteria introduced in Section \ref{sec-mcd}: MPS and EDS. This experiment evaluates how the choice of thresholds $\tau_{\mathrm{MPS}}$ and $\tau_{\mathrm{EDS}}$ affects the detection performance and demonstrates that the taxonomy can be grounded in measurable and stable statistical properties.
\begin{figure}[t]
    \centering
    \includegraphics[width=\linewidth]{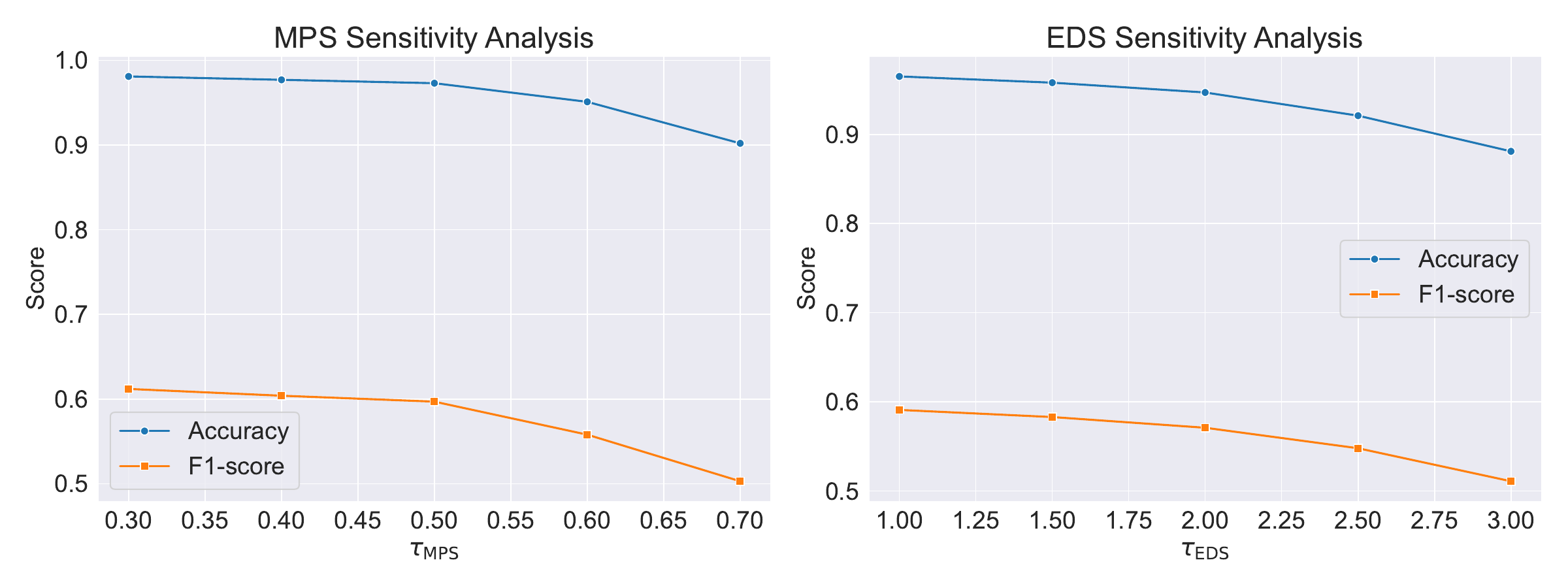}
    \caption{Sensitivity analysis of the proposed maliciousness criteria.}
    \label{fig-mps-eds}
\end{figure}

Figure \ref{fig-mps-eds} presents the performance of Concept QuickLook under varying thresholds for MPS and EDS. For MPS, which detects concepts that are intrinsically malicious, the accuracy and F1-score remain stable when the threshold is in a reasonable range. As the threshold increases, the system becomes more conservative and gradually loses sensitivity to disguised harmful prototypes, resulting in a predictable performance decay. This stability indicates that MPS effectively captures proximity to harmful semantic prototypes in the embedding space. For EDS, used to detect mislabeled or intentionally mismatched concepts, we observe a similar trend. Moderate thresholds yield the best balance between rejecting maliciously mislabeled concepts and retaining legitimate class members. Extremely low thresholds introduce excessive false positives, while overly high thresholds allow mislabeled concepts to pass. The smooth trend illustrates that the embedding distribution exhibits clear class-level structure, and deviations caused by malicious mismatches are reliably measurable.

\subsubsection{Adversarial Evasion Robustness}
\begin{figure}[t]
	\centering
	\includegraphics[width=\linewidth]{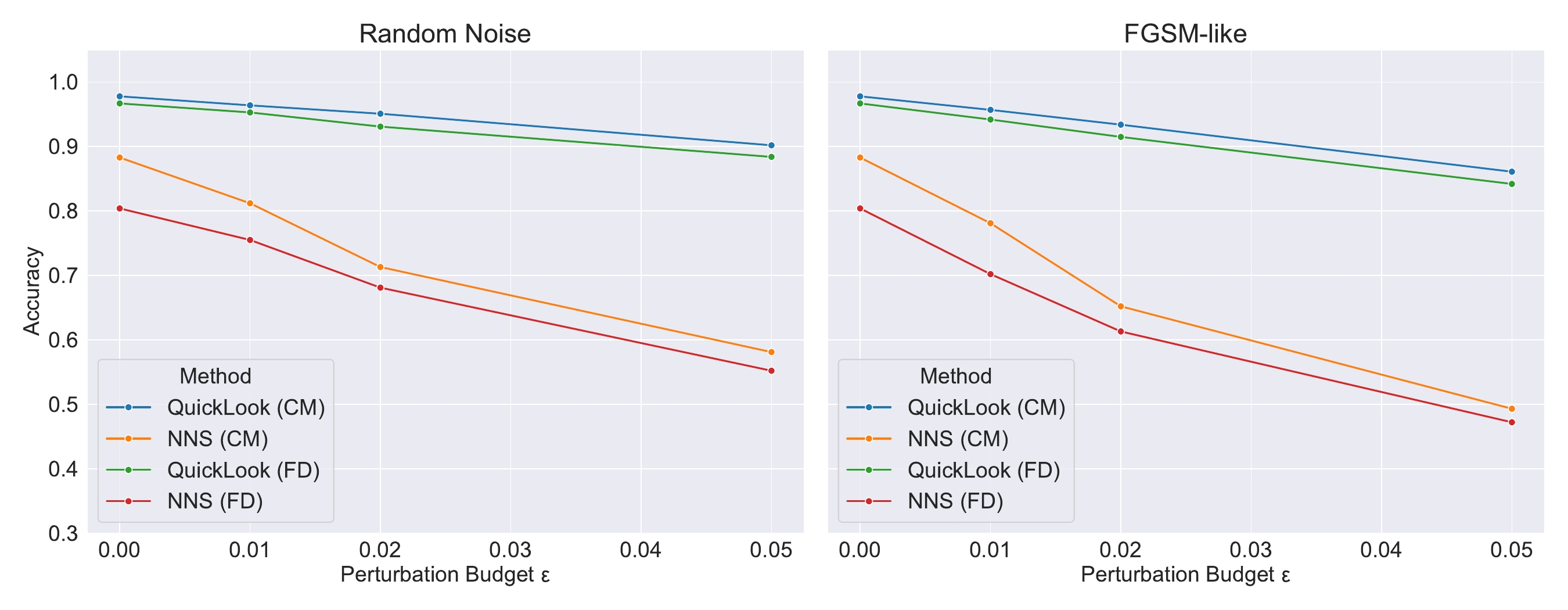}
	\caption{Adversarial evasion robustness under embedding-level perturbations.}
	\label{figure-adv}
\end{figure}

To evaluate the resilience of Concept QuickLook under adversarial evasion, we simulate embedding-level perturbations consistent with the extended threat model in Section \ref{sec-tm}. Given a embedding vector $e$, we construct an adversarial example $e' = e + \delta,\quad \|\delta\|_2 \le \epsilon$, and assess the stability of both Concept Matching and Fuzzy Detection. We conduct two types of perturbation: (\romannumeral 1) random noise with controlled magnitude, and (\romannumeral 2) gradient-guided perturbations generated via a one-step FGSM-like update. These two perturbation types serve complementary purposes: random noise evaluates local stability under non-adaptive embedding perturbations, while the FGSM-like update provides a simple first-order adversarial stress test of whether the detector can be pushed across a decision boundary by a directed perturbation. We evaluate Concept Matching (CM) and Fuzzy Detection (FD) for both Concept QuickLook and NNS under random noise and FGSM-like perturbations with varying budgets $\epsilon$. We report accuracy, F1-score, and FDS under varying perturbation budgets $\epsilon$, as shown in the Figure \ref{figure-adv}.





The results show that both detection modes remain stable under small perturbations and exhibit gradual degradation as $\epsilon$ increases. This suggests that the semantic decision of the detector is not overly sensitive to minor local variations in the embedding space. NNS baselines degrade more rapidly, whereas Concept QuickLook remains comparatively more stable because its margin-based training encourages class-wise separation rather than relying only on the nearest stored embedding. From a deployment perspective, these results indicate that QuickLook is reasonably resilient to low-budget embedding-level evasion, although its robustness weakens as perturbations become larger.

\subsubsection{Number of Concept Embedding Vectors}
A concept corresponds to a concept file which usually has one or more embedding vectors. For the tasks of extracting concepts and generating images using the concepts, the number of embedding vectors devoted to the production of the concepts will affect the effectiveness of the extraction and the quality of the generated images. In practice, although there is no absolute limit to the number of concept embedding vectors, the more embedding vectors there are, the storage space occupied by the corresponding concept file becomes larger. In addition, the marginal effect of increasing the number of embedding vectors on the quality of generation becomes more apparent. Therefore, we investigate how the number of embedding vectors affects the malicious concept detection performance of the QuickLook model. Since the sharing platform only provides concept files with a specific number of embedding vectors for download (the number of embedding vectors in a concept file is not fixed). For the convenience of conducting experiments and analyzing the experimental results, we adopted a method of extracting the same concepts ourselves, but with different quantities of embedding vectors.
\begin{figure*}[t]
	\centering
	\includegraphics[width=\linewidth]{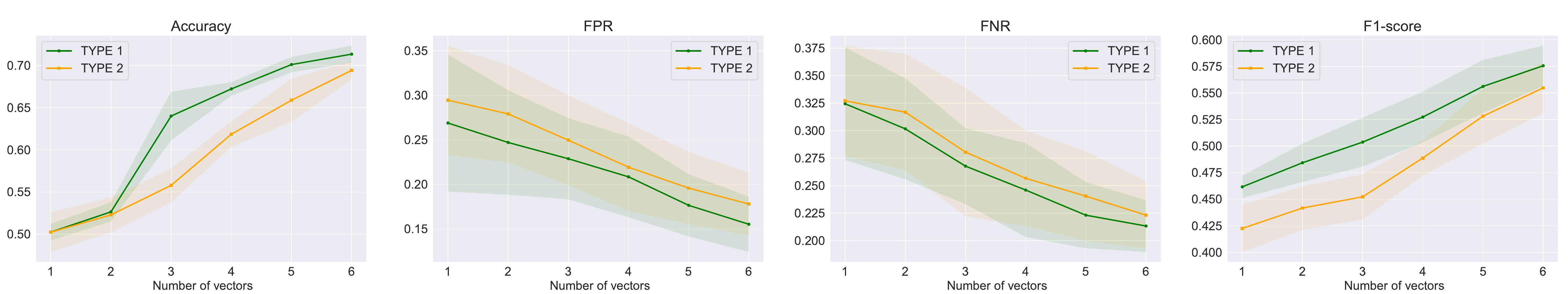}
	\caption{Performance of the QuickLook model for different numbers of concept embedding vectors.}
	\label{figure10-vectors}
\end{figure*}

We designed an experiment to verify the impact of the number of concept embedding vectors on the performance of our QuickLook model in detecting malicious concepts. Specifically, for a certain concept, we extracted it six times, with the number of embedding vectors increasing by one each time. That is, for the same concept extracted, the number of embedding vectors in the concept file ranges from 1 to 6. The value of $n$ for $e^c_n$ is $n=\{1,2,3,4,5,6\}$. We evaluated the detection performance (unseen data) of our QuickLook model for different numbers of concept embedding vectors through robustness experiments, using multiple metrics including Accuracy, FPR, FNR, and F1-score. The experimental results are illustrated in Figure \ref{figure10-vectors}. As shown in the figure, overall, the detection performance of our QuickLook model improves as the number of concept embedding vectors increases. This phenomenon is reflected across all metrics.

\subsubsection{Stable Diffusion Model Versions}
Concept image generation not only requires concept files and their corresponding pseudo-words as one of the personalized generation conditions, but also relies on large-scale T2I generation models such as SD. In practice, the input to SD consists of a prompt composed of text combined with pseudo-words. We noticed that the more common versions used in the concept sharing platform are SD1.5 and SD2.0. Therefore, we conducted evaluation experiments on the QuickLook model using the same version of SD. In addition, concepts extracted based on SD1.5 and SD2.0 have different lengths of embedding vectors, resulting in the concepts from different versions of SD not being interchangeable. SD1.5 and SD2.0 correspond to completely different embedding spaces. Let $e_{\rm{V1.5}}$ and $e_{\rm{V2.0}}$ respectively denote two different embedding spaces, and $e_{\rm{V1.5}}, e_{\rm{V2.0}} \in \mathcal{E}$.
\begin{figure}[t]
	\centering
	\includegraphics[width=\linewidth]{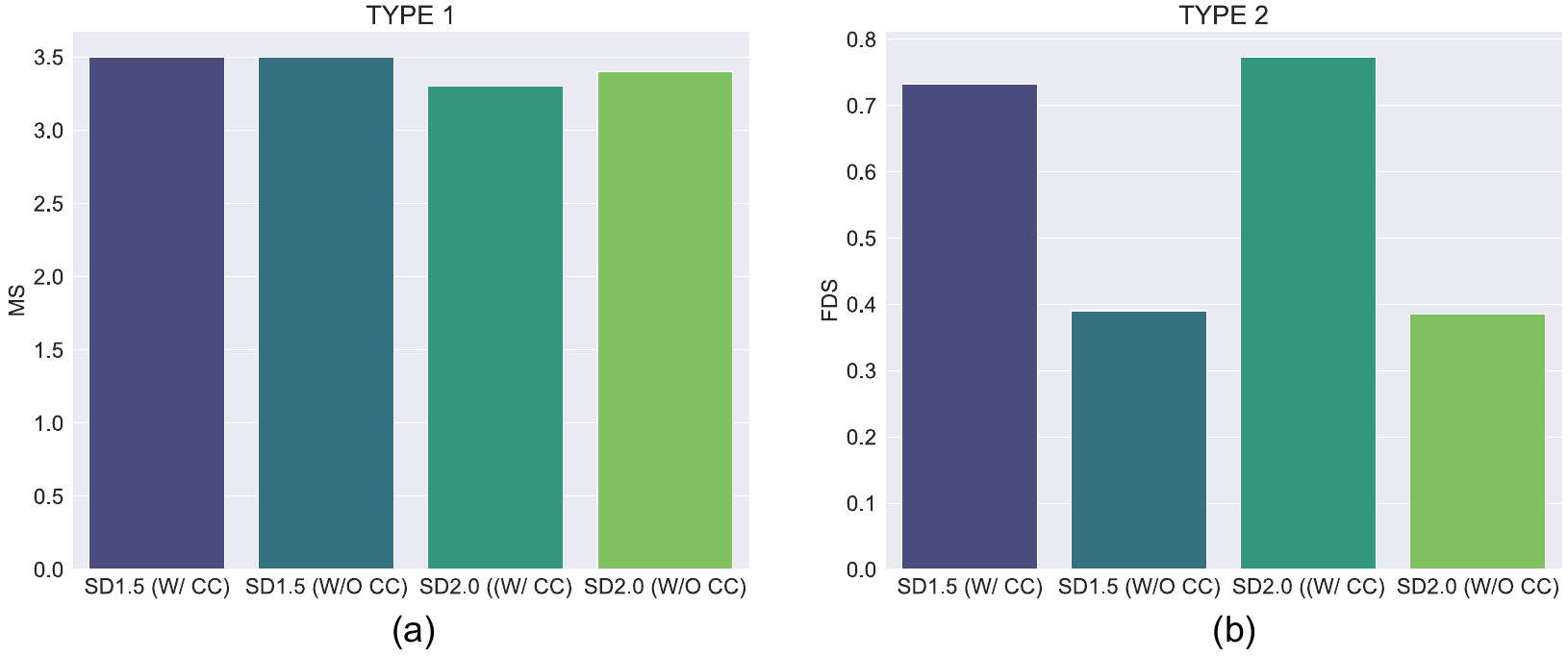}
	\caption{Performance of the QuickLook model for Stable Diffusion model versions. (a) is a comparative evaluation of SD versions for TYPE 1, and (b) is for TYPE 2.}
	\label{figure11-versions}
\end{figure}

To address the issue of different versions of the SD model, we designed a robust experiment to verify the detection performance of the QuickLook model. On one hand, for the purpose of evaluation and comparison, we standardized the number of concept embedding vectors to 6. On the other hand, we conducted comparative experiments with different versions of SD for TYPE 1 and TYPE 2, respectively, and also conducted experimental comparisons for consistent and inconsistent concept classes (class consistency). As shown in Figure \ref{figure11-versions}, the detection performance (unseen data) of the QuickLook model for different SD versions. It is clear that the fuzzy detection function of the QuickLook model performs similarly in different versions of SD and does not depend on the class of concepts. For the concept matching function, the QuickLook model performs similarly across different versions, with FDS between the same concept classes exceeding 0.7; FDS between different versions and different same concept classes are similar and below 0.4.

\subsubsection{User-Facing Result Presentation and Human-in-the-Loop Moderation}
To enhance the practical usability of Concept QuickLook, we further outline how detection outcomes can be presented to platform moderators. Both operational modes produce normalized confidence or risk scores (as detailed in Section \ref{sec-ucre1} and Section \ref{sec-ucre2}), which can be directly mapped to user-facing indicators. In deployment, platforms may display these results through simple visual cues, such as color-coded risk levels or warning banners that highlight potentially harmful or ambiguous concepts.

High-risk cases naturally receive immediate attention, while medium- and low-risk concepts can be scheduled for incremental human review. This design enables a human-in-the-loop moderation workflow, where the system performs large-scale automatic screening and human moderators focus on the most uncertain or safety-critical samples. Because Concept QuickLook operates without image generation, these signals can be computed and surfaced with minimal latency, supporting real-time decision-making during concept uploads. Such interpretability and structured prioritization make the framework actionable for integration into existing moderation pipelines.

\subsubsection{Hyperparameter Tuning and Sensitivity}
The margin parameter $\kappa$ is introduced during model training (see Section \ref{sec-qmi}) to enforce a minimum separation between embeddings of different classes. This shaping of the embedding geometry influences downstream detection, because both Concept Matching and Fuzzy Detection operate on the learned embeddings. By contrast, the threshold $\delta$ is used at inference time within the Fuzzy Detection decision rule, and it directly determines whether an unknown concept lies within the semantic neighborhood of a confirmed concept set.

We tune $\kappa$ during training by grid search on the validation set over a predefined range (for example, 0.3 to 0.7), selecting the value that maximizes validation accuracy of class-level embedding discrimination. We derive $\delta$ from the empirical distribution of intra-class distances among benign concepts. Concretely, we compute the 95-th percentile of intra-class distances on the validation set and adopt that value as a default threshold.

\begin{table*}[t]
\centering
\caption{Runtime comparison under concept generation and Concept QuickLook}
\label{tab:runtime}
\scalebox{1}{
\begin{tabular}{|l|l|c|c|l|}
\hline
Method & Type & Avg. Time & GPU Memory & Remarks \\
\hline

\multirow{2}{*}{Concept Generation} 
& Concept Matching* & 15 s & 5.3 GB & 200 denoising steps \\
& Fuzzy Detection* & 30 s & 5.3 GB & Double generations required \\
\hline

\multirow{2}{*}{Concept QuickLook} 
& Concept Matching & 1.5 s & 0.8 GB & Single forward pass \\
& Fuzzy Detection & 1.5 s & 0.8 GB & Probability-based inference \\
\hline

\end{tabular}
}
\end{table*}

Although $\kappa$ is not part of the Fuzzy Detection rule, it indirectly affects detection performance by determining how tightly classes are clustered in the embedding space. In contrast, $\delta$ directly controls acceptance or rejection at inference time. We recommend reporting a sensitivity analysis. A practical protocol is to vary each parameter by plus or minus twenty percent around its chosen value and measure changes in accuracy and F1. In our development experiments, small perturbations within this range produced only minor performance changes, indicating robustness. Finally, the Unseen results reported in Table \ref{tab-CWB1} and Table \ref{tab-CWB2} provide empirical evidence that the combination of a globally trained embedding (with an appropriate $\kappa$) and a data driven threshold $\delta$ generalizes to concept domains not seen during training.

\subsection{Efficiency Analysis and Runtime Comparison}
A key motivation behind Concept QuickLook is its computational efficiency compared with concept generation based inspection. Concept generation requires running a full denoising process to synthesize images before any semantic assessment can be performed. In contrast, Concept QuickLook operates directly in the embedding space, which removes the need for iterative sampling. To quantify this advantage, we conduct a small scale runtime comparison between concept generation based inspection and our embedding level inference.

Table \ref{tab:runtime} provides a detailed comparison across the two detection modes. For completeness, all reported average runtimes include model initialization as well as inference time. As shown in the table, implementing Concept Matching with concept generation requires a complete image synthesis process. For Fuzzy Detection, the computational cost is doubled because the system must generate two images, one for the unknown concept and one for the confirmed reference concept. These additional burdens are marked with an asterisk (*) in the table to indicate that the corresponding detection functionality is achieved only through repeated concept generation. In contrast, Concept QuickLook executes both Concept Matching and Fuzzy Detection entirely through a lightweight forward pass in the embedding space. This eliminates iterative denoising and results in substantial reductions in latency and GPU memory consumption. Concept QuickLook requires approximately 1.5 seconds per concept, in comparison with up to 30 seconds for concept generation based fuzzy detection. GPU memory usage is similarly reduced, with 0.8 GB required for Concept QuickLook and 5.3 GB required for concept generation based methods. These results indicate that embedding level analysis is significantly more efficient and more scalable.

\section{Future Research Direction}
With the comprehensive and in-depth exploration of concept generation and malicious concept detection research, we have gradually encountered some new situations, some of which may exceed the scope set out in this paper, but may provide inspiration for future research.

Situation \romannumeral 1: \textit{Multi-concept.} The scenario where the concept file contains multi-concept. The subject of this paper is a concept file corresponding to only one concept. However, in reality, there are cases where a concept file stores multiple sets of concepts, each using different pseudo-words. Our perspective is that it is feasible to divide the multi-concept into multiple individual concepts, allowing for detection on each concept separately.

Situation \romannumeral 2: \textit{Concept video.} There is currently a trend where text-to-video (T2V) \cite{singer2022make, Wu_2023_ICCV} generation is increasingly showing potential to surpass T2I generation. In the future, generating personalized concept videos will become possible. Consequently, \textit{malicious concept video} generation will also be inevitable, necessitating significant research attention.

Situation \romannumeral 3: \textit{Plug-and-Play.} We mentioned that both the extraction of concepts and the use of concepts for personalized image generation rely on the SD model, with different versions of the SD model corresponding to different concept embedding spaces. Therefore, in the experimental process of this paper, it is necessary to design the use of different versions of SD. Future work may focus on achieving truly plug-and-play malicious concept detection from the perspective of SD-version-free. Moreover, we plan to integrate the proposed Concept QuickLook framework into real-world concept-sharing platforms to further validate its scalability and real-time detection capability under operational conditions.

Situation \romannumeral 4: \textit{Adversarial Defense Integration.} Another promising direction is to integrate adversarial defense techniques, such as watermarking and tamper-proof provenance tracking, into the malicious concept detection framework. Combining proactive watermark-based protection with reactive detection methods like Concept QuickLook could form a more comprehensive AIGC security defense ecosystem.

\section{Conclusion}
We present Concept QuickLook, a malicious concept detection framework for the concept sharing process. This work formally defines malicious concepts and introduces two targeted operational modes, Concept Matching and Fuzzy Detection, which jointly address the challenges posed by inherently harmful concepts and by mislabeled or inconsistent submissions. Experimental results demonstrate that Concept QuickLook effectively identifies both types of malicious concepts without requiring any image generation, enabling efficient and low-overhead screening.

Beyond experimental validation, the framework is well suited for practical deployment. Since Concept QuickLook operates directly in the embedding space, it can be integrated into existing concept-sharing platforms as a lightweight pre-upload moderation module. In practice, platforms may adopt it through API-based screening, where submitted concept files are automatically analyzed before release. The model-agnostic design further enables seamless incorporation into offline auditing workflows or continuous content safety pipelines. We hope that this work provides a foundational step toward improving security awareness and mitigating the emerging risks associated with concept sharing in AIGC ecosystems.

%
%


\ifCLASSOPTIONcaptionsoff
  \newpage
\fi



%

\begin{thebibliography}{1}
\bibliographystyle{IEEEtran}


\bibitem{sdpaper}
R.~Rombach, A.~Blattmann, D.~Lorenz, P.~Esser, and B.~Ommer, ``High-resolution image synthesis with latent diffusion models,'' in \emph{Proc. IEEE Conf. Comput. Vis. Pattern Recognit.}, 2022, pp. 10\,674--10\,685.

\bibitem{ramesh2022}
A.~Ramesh, P.~Dhariwal, A.~Nichol, C.~Chu, and M.~Chen, ``Hierarchical text-conditional image generation with clip latents,'' \emph{arXiv:2204.06125}, 2022.

\bibitem{NEURIPS2022_ec795aea}
C.~Saharia, W.~Chan, S.~Saxena, L.~Li, J.~Whang, E.~L. Denton, K.~Ghasemipour, R.~Gontijo~Lopes, B.~Karagol~Ayan, T.~Salimans, J.~Ho, D.~J. Fleet, and M.~Norouzi, ``Photorealistic text-to-image diffusion models with deep language understanding,'' in \emph{Proc. Adv. Neural Inform. Process. Syst.}, vol.~35, 2022, pp. 36\,479-36\,494.

\bibitem{bar2023multidiffusion}
O.~Bar-Tal, L.~Yariv, Y.~Lipman, and T.~Dekel, ``Multidiffusion: Fusing diffusion paths for controlled image generation,'' in \emph{Proc. Int. Conf. Mach. Learn.}, 2023, pp. 1737-1752.

\bibitem{10841434}
S.~Pang, Y.~Rao, Z.~Lu, H.~Wang, Y.~Zhou, and M.~Xue, ``Pridm: Effective and universal private data recovery via diffusion models,'' \emph{IEEE Trans. on Dependable and Secure Comput.}, pp. 1--17, 2025.

\bibitem{ma2023subject}
J.~Ma, J.~Liang, C.~Chen, and H.~Lu, ``Subject-diffusion: Open domain personalized text-to-image generation without test-time fine-tuning,'' in \emph{Proc. ACM SIGGRAPH Conf. Pap.}, 2024.

\bibitem{shi2023instantbooth}
J.~Shi, W.~Xiong, Z.~Lin, and H.~J. Jung, ``Instantbooth: Personalized text-to-image generation without test-time finetuning,'' in \emph{Proc. IEEE Conf. Comput. Vis. Pattern Recognit.}, 2024, pp. 8543--8552.

\bibitem{gal2022image}
R.~Gal, Y.~Alaluf, Y.~Atzmon, O.~Patashnik, A.~H. Bermano, G.~Chechik, and D.~Cohen-Or, ``An image is worth one word: Personalizing text-to-image generation using textual inversion,'' in \emph{Proc.	Int. Conf. Learn. Represent.}, 2022.

\bibitem{Ruiz_2023_CVPR}
N.~Ruiz, Y.~Li, V.~Jampani, Y.~Pritch, M.~Rubinstein, and K.~Aberman, ``Dreambooth: Fine tuning text-to-image diffusion models for subject-driven generation,'' in \emph{Proc. IEEE Conf. Comput. Vis. Pattern Recognit.}, 2023, pp. 22\,500--22\,510.

\bibitem{Kumari_2023_CVPR}
N.~Kumari, B.~Zhang, R.~Zhang, E.~Shechtman, and J.-Y. Zhu, ``Multi-concept customization of text-to-image diffusion,'' in \emph{Proc. IEEE Conf. Comput. Vis. Pattern Recognit.}, 2023, pp. 1931-1941.

\bibitem{liu2023cones}
Z.~Liu, R.~Feng, K.~Zhu, Y.~Zhang, K.~Zheng, Y.~Liu, D.~Zhao, J.~Zhou, and Y.~Cao, ``Cones: Concept neurons in diffusion models for customized generation,'' in \emph{Proc. Int. Conf. Mach. Learn.}, 2023, pp. 21\,548-21\,566.

\bibitem{liu2023cones2}
Z.~Liu, Y.~Zhang, Y.~Shen, K.~Zheng, K.~Zhu, R.~Feng, Y.~Liu, D.~Zhao, J.~Zhou, and Y.~Cao, ``Cones 2: customizable image synthesis with multiple subjects,'' in \emph{Proc. Adv. Neural Inform. Process. Syst.}, vol.~37, 2024, pp. 57\,500-57\,519.

\bibitem{li2024blip}
D.~Li, J.~Li, and S.~Hoi, ``Blip-diffusion: Pre-trained subject representation for controllable text-to-image generation and editing,'' in \emph{Proc. Adv. Neural Inform. Process. Syst.}, vol.~36, 2023, pp. 30\,146--30\,166.

\bibitem{10.1145/3659578}
E.~Richardson, K.~Goldberg, Y.~Alaluf, and D.~Cohen-Or, ``ConceptLab: Creative concept generation using VLM-guided diffusion prior constraints,'' \emph{ACM Trans. Graph.}, vol.~43, no.~3, 2024.


\bibitem{10489849}
G.~Sun, W.~Liang, J.~Dong, J.~Li, Z.~Ding, and Y.~Cong, ``Create your world: Lifelong text-to-image diffusion,'' \emph{IEEE Trans. Pattern Anal. Mach. Intell.}, vol.~46, no.~9, pp. 6454-6470, 2024.

\bibitem{safaee2023clic}
M.~Safaee, A.~Mikaeili, O.~Patashnik, D.~Cohen-Or, and A.~Mahdavi-Amiri, ``Clic: Concept learning in context,'' in \emph{Proc. IEEE Conf. Comput. Vis. Pattern Recognit.}, 2024, pp. 6924-6933.

\bibitem{Gandikota_2024_WACV}
R.~Gandikota, H.~Orgad, Y.~Belinkov, J.~Materzy\'nska, and D.~Bau, ``Unified concept editing in diffusion models,'' in \emph{Proc. IEEE Winter Conf. Appl. Comput. Vis.}, 2024, pp. 5111-5120.

\bibitem{Van_Le_2023_ICCV}
T.~Van~Le, H.~Phung, T.~H. Nguyen, Q.~Dao, N.~N. Tran, and A.~Tran, ``Anti-dreambooth: Protecting users from personalized text-to-image synthesis,'' in \emph{Proc. Int. Conf. Comput. Vis.}, 2023, pp. 2116-2127.

\bibitem{zhang2023backdooring}
Y.~Wu, J.~Zhang, F.~Kerschbaum, and T.~Zhang, ``Backdooring textual inversion for concept censorship,'' \emph{arXiv:2308.10718}, 2023.

\bibitem{Kumari_2023_ICCV}
N.~Kumari, B.~Zhang, S.-Y. Wang, E.~Shechtman, R.~Zhang, and J.-Y. Zhu, ``Ablating concepts in text-to-image diffusion models,'' in \emph{Proc. Int. Conf. Comput. Vis.}, 2023, pp. 22\,691--22\,702.

\bibitem{lyu2023onedimensional}
M.~Lyu, Y.~Yang, H.~Hong, H.~Chen, X.~Jin, Y.~He, H.~Xue, J.~Han, and G.~Ding, ``One-dimensional adapter to rule them all: Concepts, diffusion models and erasing applications,'' in \emph{Proc. IEEE Conf. Comput. Vis. Pattern Recognit.}, 2024, pp. 7559-7568.

\bibitem{feng2023catch}
W.~Feng, J.~He, J.~Zhang, T.~Zhang, W.~Zhou, W.~Zhang, and N.~Yu, ``Catch you everything everywhere: Guarding textual inversion via concept watermarking,'' \emph{arXiv:2309.05940}, 2023.

\bibitem{Degeneration-Tuning}
Z.~Ni, L.~Wei, J.~Li, S.~Tang, Y.~Zhuang, and Q.~Tian, ``Degeneration-tuning: Using scrambled grid shield unwanted concepts from stable diffusion,'' in \emph{Proc. ACM Int. Conf. Multimedia}, 2023, p. 8900–8909.

\bibitem{tsai2023ring}
Y.-L. Tsai, C.-Y. Hsu, C.~Xie, C.-H. Lin, J.~Y. Chen, B.~Li, P.-Y. Chen, C.-M. Yu, and C.-Y. Huang, ``Ring-a-bell! how reliable are concept removal methods for diffusion models?'' in \emph{Proc. Int. Conf. Learn. Represent.}, 2024.

\bibitem{11316185}
K.~Gao, Y.~Zhu, Y.~Li, J.~Bai, Y.~Yang, Z.~Li, and S.-T. Xia, ``Toward dataset copyright evasion attack against personalized text-to-image diffusion models,'' \emph{IEEE Trans. Inf. Forensics Secur.}, vol.~21, pp. 725--740, 2026.

\bibitem{kim2025trainingfree}
M.~Kim, D.~Kim, A.~Yusuf, S.~Ermon, and M.~Park, ``Training-free safe denoisers for safe use of diffusion models,'' in \emph{Proc. Adv. Neural Inform. Process. Syst.}, 2025.

\bibitem{11395536}
Y.~Wang, Z.~Huang, Z.~Su, A.~Prugel-Bennett, and X.~Hong, ``Penny-wise and pound-foolish in ai-generated image detection,'' \emph{IEEE Trans. Pattern Anal. Mach. Intell.}, pp. 1--14, 2026.

\bibitem{cao2023comprehensive}
Y.~Cao, S.~Li, Y.~Liu, Z.~Yan, Y.~Dai, P.~S. Yu, and L.~Sun, ``A comprehensive survey of ai-generated content (aigc): A history of generative ai from gan to chatgpt,'' \emph{arXiv:2303.04226}, 2023.

\bibitem{wang2023security}
T.~Wang, Y.~Zhang, S.~Qi, R.~Zhao, Z.~Xia, and J.~Weng, ``Security and privacy on generative data in aigc: A survey,'' \emph{ACM Comput. Surv.}, 2024.

\bibitem{xu2024unleashing}
M.~Xu, H.~Du, D.~Niyato, J.~Kang, Z.~Xiong, S.~Mao, Z.~Han, A.~Jamalipour, D.~I. Kim, X.~Shen \emph{et~al.}, ``Unleashing the power of edge-cloud generative ai in mobile networks: A survey of aigc services,'' \emph{IEEE Commun. Surv. Tutorials}, 2024.

\bibitem{10230895}
F.~Zhan, Y.~Yu, R.~Wu, J.~Zhang, S.~Lu, L.~Liu, A.~Kortylewski, C.~Theobalt, and E.~Xing, ``Multimodal image synthesis and editing: The generative ai era,'' \emph{IEEE Trans. Pattern Anal. Mach. Intell.}, vol.~45, no.~12, pp. 15\,098-15\,119, 2023.

\bibitem{goodfellow2014generative}
I.~Goodfellow, J.~Pouget-Abadie, M.~Mirza, B.~Xu, D.~Warde-Farley, S.~Ozair, A.~Courville, and Y.~Bengio, ``Generative adversarial nets,'' \emph{Proc. Adv. Neural Inform. Process. Syst.}, vol.~27, 2014.

\bibitem{10.1145/3439723}
Z.~Wang, Q.~She, and T.~E. Ward, ``Generative adversarial networks in computer vision: A survey and taxonomy,'' \emph{ACM Comput. Surv.}, vol.~54, no.~2, 2021.

\bibitem{reed2016generative}
S.~Reed, Z.~Akata, X.~Yan, L.~Logeswaran, B.~Schiele, and H.~Lee, ``Generative adversarial text to image synthesis,'' in \emph{Proc. Int. Conf. Mach. Learn.}, 2016, pp. 1060-1069.

\bibitem{Cheng9156682}
J.~Cheng, F.~Wu, Y.~Tian, L.~Wang, and D.~Tao, ``Rifegan: Rich feature generation for text-to-image synthesis from prior knowledge,'' in \emph{Proc. IEEE Conf. Comput. Vis. Pattern Recognit.}, 2020, pp. 10\,908--10\,917.

\bibitem{Huang2021Unifying}
Y.~Huang, H.~Xue, B.~Liu, and Y.~Lu, ``Unifying multimodal transformer for bi-directional image and text generation,'' in \emph{Proc. ACM Int. Conf. Multimedia}, 2021, p. 1138–1147.

\bibitem{Ruan9710042}
S.~Ruan, Y.~Zhang, K.~Zhang, Y.~Fan, F.~Tang, Q.~Liu, and E.~Chen, ``Dae-gan: Dynamic aspect-aware gan for text-to-image synthesis,'' in \emph{Proc. Int. Conf. Comput. Vis.}, 2021, pp. 13\,940--13\,949.

\bibitem{pmlr-v37-sohl-dickstein15}
J.~Sohl-Dickstein, E.~Weiss, N.~Maheswaranathan, and S.~Ganguli, ``Deep unsupervised learning using nonequilibrium thermodynamics,'' in \emph{Proc. Int. Conf. Mach. Learn.}, 2015, pp. 2256-2265.

\bibitem{NEURIPS2020_4c5bcfec}
J.~Ho, A.~Jain, and P.~Abbeel, ``Denoising diffusion probabilistic models,'' in \emph{Proc. Adv. Neural Inform. Process. Syst.}, vol.~33, 2020, pp. 6840--6851.

\bibitem{Croitoru10081412}
F.-A. Croitoru, V.~Hondru, R.~T. Ionescu, and M.~Shah, ``Diffusion models in vision: A survey,'' \emph{IEEE Trans. Pattern Anal. Mach. Intell.}, vol.~45, no.~9, pp. 10\,850--10\,869, 2023.

\bibitem{Yang2023Diffusion}
L.~Yang, Z.~Zhang, Y.~Song, S.~Hong, R.~Xu, Y.~Zhao, W.~Zhang, B.~Cui, and M.-H. Yang, ``Diffusion models: A comprehensive survey of methods and applications,'' \emph{ACM Comput. Surv.}, vol.~56, no.~4, 2023.

\bibitem{NEURIPS2021_49ad23d1}
P.~Dhariwal and A.~Nichol, ``Diffusion models beat gans on image synthesis,'' in \emph{Proc. Adv. Neural Inform. Process. Syst.}, vol.~34, 2021, pp. 8780--8794.

\bibitem{Gu9879180}
S.~Gu, D.~Chen, J.~Bao, F.~Wen, B.~Zhang, D.~Chen, L.~Yuan, and B.~Guo, ``Vector quantized diffusion model for text-to-image synthesis,'' in \emph{Proc. IEEE Conf. Comput. Vis. Pattern Recognit.}, 2022, pp. 10\,686--10\,696.

\bibitem{liu2024instaflow}
X.~Liu, X.~Zhang, J.~Ma, J.~Peng, and qiang liu, ``Instaflow: One step is enough for high-quality diffusion-based text-to-image generation,'' in \emph{Proc. Int. Conf. Learn. Represent.}, 2024.

\bibitem{zhang2023hive}
S.~Zhang, X.~Yang, Y.~Feng, C.~Qin, C.-C. Chen, N.~Yu, Z.~Chen, H.~Wang, S.~Savarese, S.~Ermon, C.~Xiong, and R.~Xu, ``Hive: Harnessing human feedback for instructional visual editing,'' in \emph{Proc. IEEE Conf. Comput. Vis. Pattern Recognit.}, 2024, pp. 9026--9036.

\bibitem{NEURIPS2023_f8ad010c}
K.~Huang, K.~Sun, E.~Xie, Z.~Li, and X.~Liu, ``T2i-compbench: A comprehensive benchmark for open-world compositional text-to-image generation,'' in \emph{Proc. Adv. Neural Inform. Process. Syst.}, vol.~36, 2023, pp. 78\,723-78\,747.

\bibitem{hu2022lora}
E.~J. Hu, Y.~Shen, P.~Wallis, Z.~Allen-Zhu, Y.~Li, S.~Wang, L.~Wang, and W.~Chen, ``Lo{RA}: Low-rank adaptation of large language models,'' in \emph{Proc. Int. Conf. Learn. Represent.}, 2022.

\bibitem{10.1145/3618315}
Y.~Vinker, A.~Voynov, D.~Cohen-Or, and A.~Shamir, ``Concept decomposition for visual exploration and inspiration,'' \emph{ACM Trans. Graph.}, vol.~42, no.~6, 2023.

\bibitem{Avrahami2023Break}
O.~Avrahami, K.~Aberman, O.~Fried, D.~Cohen-Or, and D.~Lischinski, ``Break-a-scene: Extracting multiple concepts from a single image,'' in \emph{Proc. SIGGRAPH Asia Conf. Pap.}, 2023.

\bibitem{zhao2023catversion}
R.~Zhao, M.~Zhu, S.~Dong, N.~Wang, and X.~Gao, ``Catversion: Concatenating embeddings for diffusion-based text-to-image personalization,'' \emph{arXiv:2311.14631}, 2023.

\bibitem{NEURIPS2023_d33b177b}
K.~Sohn, L.~Jiang, J.~Barber, K.~Lee, N.~Ruiz, D.~Krishnan, H.~Chang, Y.~Li, I.~Essa, M.~Rubinstein, Y.~Hao, G.~Entis, I.~Blok, and D.~Castro~Chin, ``Styledrop: Text-to-image generation in any style,'' in \emph{Proc. Adv. Neural Inform. Process. Syst.}, vol.~36, 2023, pp. 66\,860--66\,889.

\bibitem{Zhang_2023_CVPR}
Y.~Zhang, N.~Huang, F.~Tang, H.~Huang, C.~Ma, W.~Dong, and C.~Xu, ``Inversion-based style transfer with diffusion models,'' in \emph{Proc. IEEE Conf. Comput. Vis. Pattern Recognit.}, 2023, pp. 10\,146--10\,156.

\bibitem{Lu_2023_CVPR}
H.~Lu, H.~Tunanyan, K.~Wang, S.~Navasardyan, Z.~Wang, and H.~Shi, ``Specialist diffusion: Plug-and-play sample-efficient fine-tuning of text-to-image diffusion models to learn any unseen style,'' in \emph{Proc. IEEE Conf. Comput. Vis. Pattern Recognit.}, 2023, pp. 14\,267--14\,276.

\bibitem{11112743}
S.~Wu, H.~Sun, T.~Zhu, and W.~Zhou, ``Backdoor defense for text encoders in text-to-image generative models,'' \emph{IEEE Trans. Dependable Secure Comput.}, vol.~22, no.~6, pp. 7139--7156, 2025.

\bibitem{11300728}
Z.~Wang, J.~Zhang, S.~Shan, and X.~Chen, ``Dynamic attention analysis for backdoor detection in text-to-image diffusion models,'' \emph{IEEE Trans. Pattern Anal. Mach. Intell.}, vol.~48, no.~3, pp. 3652--3665, 2026.

\bibitem{Li_2025_ICCV}
Z.~Li, H.~Qu, J.~Kuen, J.~Gu, Q.~Ke, J.~Liu, and H.~Rahmani, ``Diffip: Representation fingerprints for robust ip protection of diffusion models,'' in \emph{Proc. Int. Conf. Comput. Vis.}, 2025, pp. 17\,035--17\,045.


\bibitem{pmlr-v139-radford21a}
A.~Radford, J.~W. Kim, C.~Hallacy, A.~Ramesh, G.~Goh, S.~Agarwal, G.~Sastry, A.~Askell, P.~Mishkin, J.~Clark, G.~Krueger, and I.~Sutskever, ``Learning transferable visual models from natural language supervision,'' in \emph{Proc. Int. Conf. Mach. Learn.}, 2021, pp. 8748-8763.

\bibitem{Gore-Blood-Dataset-v1.0}
NeuralShell, ``Gore blood dataset,'' 2023.

\bibitem{gore-kfldh_dataset}
R.~Universe, ``Gore dataset,'' 2023.

\bibitem{8733051}
J.~Johnson, M.~Douze, and H.~Jégou, ``Billion-scale similarity search with gpus,'' \emph{IEEE Trans. Big Data}, vol.~7, no.~3, pp. 535--547, 2021.

\bibitem{singer2022make}
U.~Singer, A.~Polyak, T.~Hayes, X.~Yin, J.~An, S.~Zhang, Q.~Hu, H.~Yang, O.~Ashual, O.~Gafni, D.~Parikh, S.~Gupta, and Y.~Taigman, ``Make-a-video: Text-to-video generation without text-video data,'' in \emph{Proc. Int. Conf. Learn. Represent.}, 2023.

\bibitem{Wu_2023_ICCV}
J.~Z. Wu, Y.~Ge, X.~Wang, S.~W. Lei, Y.~Gu, Y.~Shi, W.~Hsu, Y.~Shan, X.~Qie, and M.~Z. Shou, ``Tune-a-video: One-shot tuning of image diffusion models for text-to-video generation,'' in \emph{Proc. Int. Conf. Comput. Vis.}, 2023, pp. 7623--7633.


\end{thebibliography}



\begin{IEEEbiography}[{\includegraphics[width=1in,height=1.25in,clip,keepaspectratio]{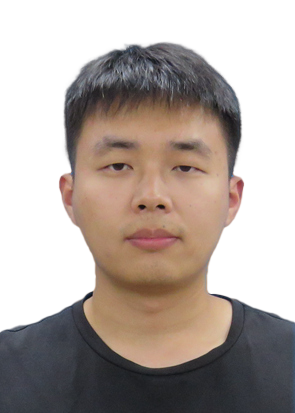}}]{Kun Xu}
received the B.E. and M.E. degrees from Anhui University of Science and Technology, Huainan, China, in 2020 and 2023, respectively. He is currently working toward the Ph.D. degree with the Nanjing University of Aeronautics and Astronautics, Nanjing, China. His research interests include generative model security, multimedia forensics and trustworthy machine learning.
\end{IEEEbiography}

\begin{IEEEbiography}[{\includegraphics[width=1in,height=1.25in,clip,keepaspectratio]{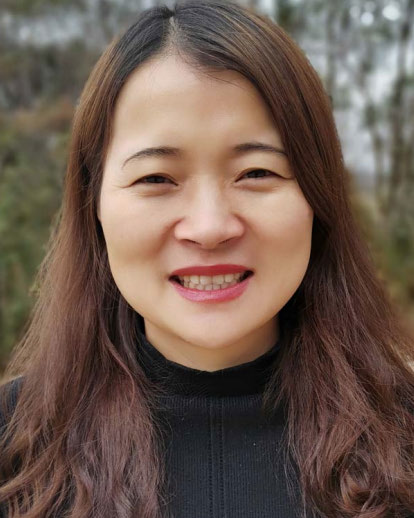}}]{Wenying Wen}
(Member, IEEE) received the M.S. degree in computational mathematics from the Inner Mongolia University of Technology, Hohhot, China, in 2010, and the Ph.D. degree in computational mathematics from Chongqing University, Chongqing, China, in 2013. She is currently a Professor with the School of Computing and Artificial Intelligence, Jiangxi University of Finance and Economics, Nanchang, China. Her research interests include image processing, multimedia security, compressive sensing security, and blockchain.
\end{IEEEbiography}



\begin{IEEEbiography}[{\includegraphics[width=1in,height=1.25in,clip,keepaspectratio]{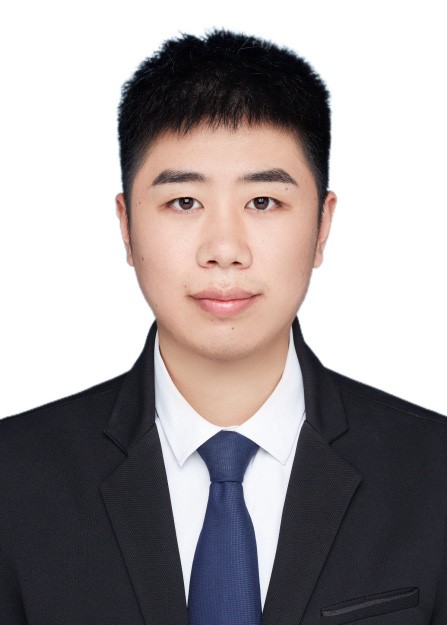}}]{Shuren Qi}
received the Ph.D. degree in computer science from the Nanjing University of Aeronautics and Astronautics, Nanjing, China, in 2024. He currently is a postdoctoral fellow at the City University of Hong Kong. He has published academic papers in top-tier venues including the ACM Computing Surveys and IEEE Transactions on Pattern Analysis and Machine Intelligence. His research involves the general topics of invariance, robustness, and explainability in computer vision, with a focus on invariant representations, for closing today's trustworthiness gap in artificial intelligence, e.g., forensic and security of visual data.
\end{IEEEbiography}


\begin{IEEEbiography}[{\includegraphics[width=1in,height=1.25in,clip,keepaspectratio]{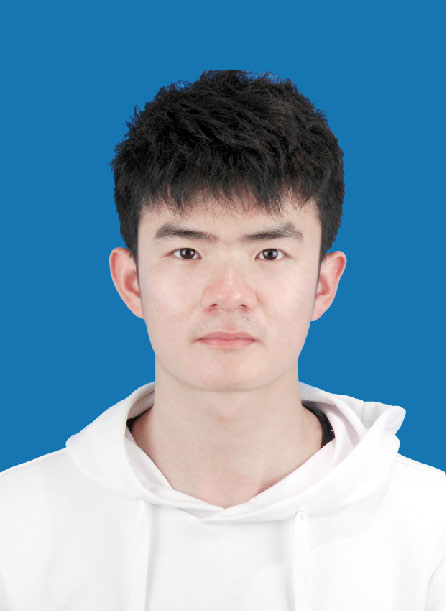}}]{Tao Wang}
received the M.S. degree in cyberspace security from the College of Computer Science and Technology, Nanjing University of Aeronautics and Astronautics, Nanjing, China, in Apr. 2024. He is currently working toward the Ph.D. degree in cyberspace security with the College of Com puter Science and Technology, Nanjing University of Aeronautics and Astronautics, Nanjing, China. His current research interest is image privacy protection.
\end{IEEEbiography}


\begin{IEEEbiography}[{\includegraphics[width=1in,height=1.25in,clip,keepaspectratio]{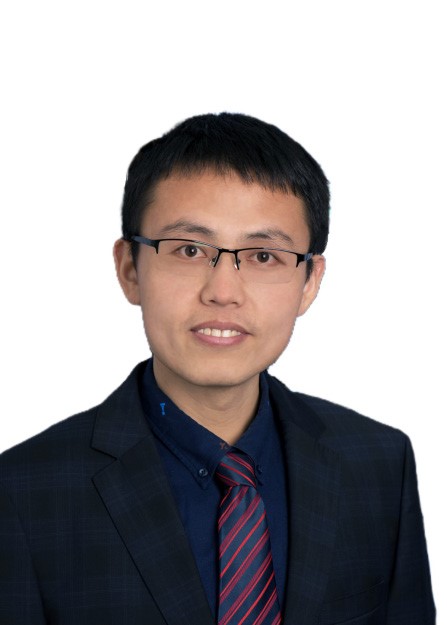}}]{Yushu Zhang}
(Senior Member, IEEE) received the B.S. degree from the School of Science, North University of China, Taiyuan, China, in 2010, and the Ph.D. degree from the College of Computer Science, Chongqing University, Chongqing, China, in 2014. He held various research positions with the City University of Hong Kong, Southwest University, the University of Macau, and Deakin University. He is currently a Professor with the College of Computer Science and Technology, Nanjing University of Aeronautics and Astronautics, Nanjing, China. His research interests include multimedia security, blockchain, and artificial intelligence. He has coauthored more than 200 refereed journal articles and conference papers in these areas. He is an Associate Editor of Information Sciences, Journal of King Saud University-Computer and Information Sciences, and Signal Processing. 
\end{IEEEbiography}


\begin{IEEEbiography}[{\includegraphics[width=1in,height=1.25in,clip,keepaspectratio]{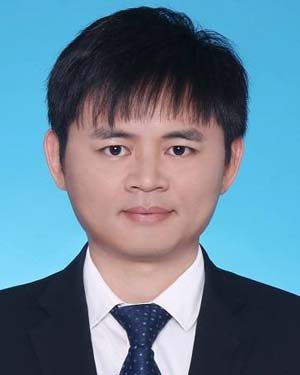}}]{Yuming Fang}
(Fellow, IEEE) received the Ph.D. degree from Nanyang Technological University, Singapore, 2013. He is currently a Professor with the School of Computing and Artificial Intelligence, Jiangxi University of Finance and Economics, Nanchang, China. His research interests include visual attention modeling, visual quality assessment, image retargeting, computer vision, 3D image/video processing. 
\end{IEEEbiography}


\vfill


\end{document}